\documentclass[11pt]{article}
\usepackage{latexsym}
\usepackage{amssymb}
\usepackage{epsf}
\usepackage{amsmath}
\usepackage[hypertex]{hyperref}
\usepackage{graphicx}

\newcommand{\gsim}{\gtrsim}

\begin{document}

\pagestyle{empty}

\begin{flushright}
SLAC-PUB-12532
\end{flushright}

\vspace{2.5cm}

\begin{center}

{\bf\LARGE Sweet Spot Supersymmetry}
\\

\vspace*{1.5cm}
{\large 
Masahiro Ibe and
Ryuichiro Kitano
} \\
\vspace*{0.5cm}

{\it Stanford Linear Accelerator Center, Stanford University,
                Stanford, CA 94309 and } \\
{\it Physics Department, Stanford University, Stanford, CA 94305}\\
\vspace*{0.5cm}

\end{center}

\vspace*{1.0cm}

\begin{abstract}
{
\baselineskip 14pt

We find that there is no supersymmetric flavor/CP problem,
$\mu$-problem, cosmological moduli/gravitino problem or dimension
four/five proton decay problem in a class of supersymmetric theories
with $O(1)$~GeV gravitino mass. The cosmic abundance of the
non-thermally produced gravitinos naturally explains the dark matter
component of the universe.  A mild hierarchy between the mass scale of
supersymmetric particles and electroweak scale is predicted, consistent
with the null result of a search for the Higgs boson at the LEP-II
experiments. A relation to the strong CP problem is addressed.  We
propose a parametrization of the model for the purpose of collider
studies.
The scalar tau lepton is the next to lightest supersymmetric particle in
a theoretically favored region of the parameter space. The lifetime of
the scalar tau is of $O(1000)$~seconds with which it is regarded as a
charged stable particle in collider experiments.
We discuss characteristic signatures and a strategy for confirmation of
this class of theories at the LHC experiments.

}
\end{abstract} 

\newpage
\baselineskip 18pt
\setcounter{page}{2}
\pagestyle{plain}
\section{Introduction}

\setcounter{footnote}{0}

In spontaneously broken supersymmetric theories, there is a spin-half
Goldstino fermion which is eaten by the gravitino as its longitudinal
components. By supersymmetry, the Goldstino must be accompanied with its
superpartner whose spin is zero if supersymmetry is broken by a vacuum
expectation value of the $F$-component of a chiral superfield. A chiral
supermultiplet is formed by the Goldstino, its scalar superpartner, and
the non-vanishing $F$-term, which we call the chiral superfield $S$.
The low energy physics is then described by matter superfields, gauge
superfields and the chiral superfield $S$.

There are variety of possibilities for couplings between matter/gauge
superfields in the supersymmetric standard model and the superfield
$S$. These possibilities have been classified as follows. If we assume
that the couplings are suppressed by the Planck scale $(M_{\rm Pl})$,
such as ${\cal L} \ni [(S / M_{\rm Pl}) W^\alpha W_\alpha]_{\theta^2}$
with $W^\alpha$ being gauge fields, the model is called the ``gravity
mediation''~\cite{Chamseddine:1982jx,Hall:1983iz}.
Another possibility that the gauge kinetic function is of the form,
${\cal L} \ni [ (\log S / (4 \pi)^2 ) W^\alpha W_\alpha ]_{\theta^2}$,
is called the ``gauge
mediation''~\cite{Dine:1981za,Dine:1993yw,Dine:1994vc,Dine:1995ag}. This
is the form we obtain after integrating out vector-like fields which obtain
masses proportional to $\langle S \rangle$~\cite{Giudice:1997ni}.  
If the coupling is more suppressed than the Planck scale, effects of the
``anomaly mediation~\cite{Randall:1998uk}'' give the largest
contribution to supersymmetry breaking terms in the Lagrangian.
Among those scenarios, gauge mediation assumes the strongest interaction
between the matter/gauge fields and $S$ while the anomaly mediation
effects are the weakest.
The size of supersymmetry breaking, $F_S$, therefore has a relation,
$F_{\rm gauge} \ll F_{\rm gravity} \ll F_{\rm anomaly}$, when we fix the
scale of gaugino/sfermion masses. The gravitino masses are $m_{3/2}^{\rm
gauge} \ll m_{3/2}^{\rm gravity} \ll m_{3/2}^{\rm anomaly}$ as $m_{3/2}
\propto F_S$.

The question is what size of the gravitino mass (i.e., the supersymmetry
breaking scale) is preferred by phenomenological and cosmological
requirements. This is an interesting question since each scenario
predicts a quite different pattern in the spectrum of the supersymmetric
particles, which we will search for at the LHC experiments.
Strategies for finding supersymmetric particles and measurements of
model parameters will be also different for different scales of
$m_{3/2}$.

There have been many model-building efforts in making supersymmetric
models realistic in each category: gravity, gauge or anomaly mediation.
If one of them had been completely successful, we could have believed in
the scenario and used the model as the {\it standard supersymmetric
model}.
However, unfortunately, there is no such a standard model so far because
of the fact that neither of these scenarios are fully realistic by
different reasons.
In the gauge and anomaly mediation scenarios, there is a problem with
the electroweak symmetry breaking, i.e., the $\mu$-problem.
The pure anomaly mediation, in addition, predicts tachyonic scalar
leptons which are not acceptable.
Although the gravity mediation scenario does not suffer from those
problems, it has been known that sizes of flavor and CP violation are
expected to be too large.
There are also cosmological constraints. In particular, in gravity
mediation models, a moduli problem caused by fields in the supersymmetry
breaking sector destroys cosmological successes of the (supersymmetric)
standard model~\cite{Coughlan:1983ci}, such as the big-bang
nucleosynthesis (BBN) and also cold dark matter by thermal-relic
neutralinos.

In this paper, we reconsider problems in supersymmetric models by using
an effective field theory described by the field $S$ and the
matter/gauge fields. By doing so, we can discuss each of these scenarios
as a different choice of functions of $S$ which define an effective
theory.
The labeling can be done by projecting the function space onto a
one-dimensional axis of the gravitino mass.
In this formulation, we find that there is a sweet spot in between the
gauge and gravity mediation ($m_{3/2} \sim O(1)$~GeV) where the theory
is perfectly consistent with various requirements.
All the classic problems, such as the flavor/CP problem and the
$\mu$-problem are absent.
The theory also avoids a cosmological moduli problem caused by the
scalar component of $S$.
Non-thermally produced gravitinos through the decay of the
$S$-condensation naturally account for dark matter of the universe.
A simple ultraviolet (UV) completion of the theory exists, which is
actually a model of grand unification without neither the
doublet-triplet splitting problem nor the proton decay problem.
Relations to the strong CP problem and the supersymmetric fine-tuning
problem are also addressed. We discuss a characteristic spectrum of
supersymmetric particles, and demonstrate how we can confirm this
scenario.

In the next section, we rewrite the various supersymmetric models in
terms of the effective Lagrangian described by the Goldstino multiplet
$S$ and particles in the minimal supersymmetric standard model (MSSM).
The section includes review of the supersymmetry breaking and its
transmission.  A concrete set-up is defined in subsection~\ref{sec:sss}
and discuss its successes there.
We then discuss low energy predictions of the framework in
Section~\ref{sec:low-energy}. A parametrization of the model and a way
of calculating the spectrum of supersymmetric particles are presented.
Collider signatures are discussed in Section~\ref{sec:lhc}. We
demonstrate a method of extracting model parameters in the case where
the scalar tau (stau) is the next to lightest supersymmetric particle
(NLSP).

\section{Theoretical set-up}

\setcounter{footnote}{0}

We construct a phenomenological Lagrangian of the supersymmetric
standard model and consider various requirements from particle physics
and cosmology.
We will arrive at a scenario with $m_{3/2} \sim 1$~GeV.

\subsection{{\boldmath $S$} sector}
\label{sec:s-sector}

We derive here a description of a supersymmetry breaking sector by the
Goldstino chiral superfield $S$. This corresponds to the construction of
the Higgs sector in the standard model. As any models of the electroweak
symmetry breaking flow into the standard model with various mass ranges
of the Higgs boson at low energy, the model below provides a standard
low energy description of a variety of supersymmetry breaking models.

We concentrate on $F$-term supersymmetry breaking scenarios as most of
the supersymmetry breaking models are of this type. To ensure a
non-vanishing vacuum expectation value of the $F$-component of a chiral
superfield $S$, we add a source term in the Lagrangian:
\begin{eqnarray}
 {\cal L} \ni m^2 F_S + {\rm h.c.}
\end{eqnarray}
This term can be expressed in terms of the superfield as follows:
\begin{eqnarray}
 W \ni m^2 S \ .
\label{eq:s-linear}
\end{eqnarray}

We can also write down an arbitrary K\"ahler potential, $K_S$, for the
kinetic and interaction terms of $S$. As long as $\partial^2 K_S / (
\partial S \partial S^\dagger ) $ is a non-singular function, $F_S \neq
0$ is obtained by the equation of motion. For example, the low energy
effective theory of the O'Raifeartaigh
model~\cite{O'Raifeartaigh:1975pr} has a K\"ahler potential:
\begin{eqnarray}
 K_S = S^\dagger S - \frac{(S^\dagger
  S)^2}{\Lambda^2} \ ,
\label{eq:s-kahler}
\end{eqnarray}
where $\Lambda$ is the mass scale of the massive fields which have been
integrated out.
In general, if $S$ carries some approximately conserving charge, the
K\"ahler potential is restricted to the form in Eq.~(\ref{eq:s-kahler}) (up
to the sign of the second term).\footnote{In fact, the cubic term in the
K\"ahler potential, $K \ni S^\dagger S^2 + {\rm h.c.}$ can be eliminated
by the shift of $S$ in general. However, once we take into account
interaction terms between $S$ and the MSSM fields, the origin of $S$ has
a definite meaning and we cannot shift away the cubic term.} The second
term gives a mass to the scalar component of $S$, $m_S$:
\begin{eqnarray}
 m_S = \frac{2 F_S}{\Lambda} = \frac{2 m^2}{\Lambda} = 2 \sqrt 3 m_{3/2}
  \left( \frac{M_{\rm Pl}}{\Lambda} \right)\ ,
\label{eq:s-mass}
\end{eqnarray}
and stabilizes the value of $S$ at
\begin{eqnarray}
 S = 0\ .
\end{eqnarray}
Here we ignored supergravity effects. The fermionic component of $S$
remains massless. This is the Goldstino fermion associated with the
spontaneous supersymmetry breaking.

Note that the existence of the chiral superfield $S$ in the above
effective theory does not necessarily mean that the supersymmetry
breaking sector contains a gauge singlet chiral superfield in the UV
theory. The $S$ field can originate from a component of some multiplets
or can be a composite operator in physics above a `cut-off' scale
$\Lambda$. It is totally a general argument that there is a gauge
singlet chiral superfield $S$ in the effective theory below the scale of
supersymmetry breaking dynamics, $\Lambda$, as long as $\Lambda^2
\gtrsim m^2$.

The Lagrangian discussed above is analogous to the Higgs sector in the
standard model. The two parameters $m^2$ and $m^2/\Lambda^2 (\sim m_S^2
/ m^2)$ correspond to the parameters $v^2$ and $\lambda_H (\sim m_h^2 /
v^2)$ in the Higgs potential, $V = (\lambda_H/4) (|H|^2 - v^2)^2$.
We should not trust this effective theory if $m^2 / \Lambda^2 \gsim
\sqrt{4 \pi}$ as it violates the unitarity of scattering amplitudes of the
gravitinos at high energy just like the standard model with $\lambda_H
\gtrsim 4 \pi$.

\subsection{Matter/gauge sector}
\label{sec:mssm}

The superpotential of the MSSM is
\begin{eqnarray}
 W_{\rm MSSM} = Q H_u U + Q \bar H_d D + L H_d E + \mu H_u H_d\ ,
\end{eqnarray}
where we suppressed the Yukawa coupling constants and flavor indices.
The last term, the $\mu$-term, is needed to give a mass to the Higgsino,
but it should not be too large. For supersymmetry to be a solution to
the hierarchy problem, i.e., $\langle H_{u,d} \rangle \ll M_{\rm Pl}$, the
$\mu$-term is necessary to be of the order of the electroweak scale (or
scale of the soft supersymmetry breaking terms).

This is called the $\mu$-problem. The fact that $\mu$ is much smaller
than the Planck scale suggests that the combination of $H_u H_d$
carries some approximately conserving charge.

There are many gauge invariant operators we can write down in addition
to the above superpotential such as
\begin{eqnarray}
 W_{R \!\!\! /} =
UDD + LLE + QLD \ ,
\label{eq:r-breaking}
\end{eqnarray}
and
\begin{eqnarray}
 W_{\rm dim. 5} = QQQL + UDUE\ .
\end{eqnarray}
These are unwanted operators as they cause too rapid proton decays.

The $\mu$-problem and the proton decay problem above are actually
related, and there is a simple solution to both problems. The
Peccei-Quinn (PQ) symmetry with the following charge assignment avoids
too large $\mu$-term and the proton decay operators.
\begin{eqnarray}
 PQ(Q) = PQ(U) = PQ(D) = PQ(L) = PQ(E) = - \frac{1}{2}\ ,
\end{eqnarray}
\begin{eqnarray}
 PQ(H_u) = PQ(H_d) = 1\ .
\end{eqnarray}
This symmetry is broken explicitly by the $\mu$-term, $PQ(\mu)=-2$.
Since it is a small breaking of the PQ symmetry, the coefficients of the
dimension five operators are sufficiently suppressed.
The unbroken $Z_4$ symmetry, which includes the $R$-parity as a
subgroup, still forbids the superpotential terms in
Eq.~(\ref{eq:r-breaking}) and ensures the stability of the lightest
supersymmetric particle (LSP), leaving us to have a candidate for dark
matter of the universe.

The Majorana neutrino mass terms, $W \ni LL H_u H_u$, are forbidden by
the PQ symmetry, but large enough coefficients can be obtained by
introducing another explicit breaking of the PQ symmetry. For example,
we can write down $LL H_u H_u/M_N$ with $PQ(M_N)=1$ without introducing
proton decay operators or too large $\mu$-term. The $Z_4$ symmetry above
is broken down to the $R$-parity with this term.

In fact, there is another symmetry which can play the same role as the
PQ symmetry, called $R$-symmetry. The charge assignment is
\begin{eqnarray}
 R(Q) = R(U) = R(D) = R(L) = R(E) = 1\ ,
\end{eqnarray}
\begin{eqnarray}
 R(H_u) = R(H_d) = 0\ .
\end{eqnarray}
Again, $R(\mu) = 2$ explicitly breaks the $R$-symmetry down to the
$R$-parity. In this case, the $LLH_u H_u$ term is allowed by the
symmetry.

In summary, there are approximate symmetries, U(1)$_{PQ}$ and U(1)$_R$,
in the Lagrangian of the MSSM.  If one of them is a good (approximate)
symmetry of the whole system, it provides us with a solution to the
$\mu$ and the proton decay problems.

\subsection{Interaction to mediate the supersymmetry breaking}
Now we discuss interaction terms between the $S$-sector and the MSSM
sector. These interactions determine the pattern of supersymmetry
breaking parameters which are relevant for low energy physics.
We review here three famous mechanisms; gravity, gauge, and anomaly
mediation models, as choices of the form of the interactions. Each of
these scenarios suffer from different problems. Understanding nature of
those problems guides us to a phenomenologically consistent model.

\subsubsection{Gravity mediation}

The simplest scenario is to assume general interaction terms suppressed
by the Planck scale. This is called the gravity mediation. The K\"ahler
potential is 
\begin{eqnarray}
 K_{\rm gravity}^{\rm (matter)} = 
-  \frac{S^\dagger S \Phi^\dagger \Phi}{M_{\rm Pl}^2}
+ \left( \frac{ S \Phi^\dagger \Phi}{M_{\rm Pl}} + {\rm
   h.c.} \right)
+ \cdots
\label{eq:gravity-matter}
\end{eqnarray}
\begin{eqnarray}
 K_{\rm gravity}^{\rm (Higgs)} &=& 
\left( H_u H_d + {\rm h.c.} \right)
+ \left( \frac{S^\dagger H_u H_d}{M_{\rm Pl}} + {\rm h.c.} \right)
+ \left( \frac{S^\dagger S  H_u H_d }{M_{\rm Pl}^2} + {\rm h.c.} \right)
\nonumber \\
&&- \frac{S^\dagger S ( H_u^\dagger H_u + H_d^\dagger H_d )}{M_{\rm
Pl}^2}
+ \cdots
\label{eq:gravity-higgs}
\end{eqnarray}
where $\Phi$ represents the quark and lepton superfields in the MSSM,
and we omit $O(1)$ coefficients.

Planck suppressed operators in the gauge kinetic function generate
gaugino masses:
\begin{eqnarray}
 f_{\rm gravity} = 
\left( {1 \over g^2} + \frac{S}{M_{\rm Pl}} \right)
W^\alpha W_\alpha\ ,
\label{eq:gravity-gauge}
\end{eqnarray}
where $g$ is the gauge coupling constant.

The first term in Eq.~(\ref{eq:gravity-matter}) and the second term in
the bracket in Eq.~(\ref{eq:gravity-gauge}) generate sfermion masses and
gaugino masses, respectively. Both of them are of $O(F_S / M_{\rm Pl})
\sim m_{3/2}$.  Therefore, the gravitino mass, $m_{3/2}$, is
$O(100)$~GeV in this scenario. The $\mu$-problem is completely solved in
a quite natural way~\cite{Giudice:1988yz}. The first and second terms in
Eq.~(\ref{eq:gravity-higgs}) generates $\mu \sim O(m_{3/2})$.

This mechanism for the $\mu$-term generation is consistent with the
discussion in the previous subsection. The $R$-symmetry introduced
before can be preserved once we assign $R(S) = 0$. The term in
Eq.~(\ref{eq:s-linear}) breaks the $R$-symmetry by $R(m^2) = 2$ at the
intermediate scale, $m^2 = \sqrt 3 m_{3/2} M_{\rm Pl}$. This is still
small enough for the proton decay operators.

On the other hand, the natural solution to the $\mu$-problem is not
compatible with the PQ symmetry. The term in
Eq.~(\ref{eq:gravity-gauge}) restricts the PQ charge of $S$ to be
vanishing, whereas the term responsible for the $\mu$-term generation,
$K \ni S^\dagger H_u H_d$, determines that $PQ(S) = 2$.
The PQ symmetry must, therefore, be maximally violated.
This implies that none of the terms in Eq.~(\ref{eq:gravity-higgs}) can
be forbidden by approximate symmetries of the theory. This fact becomes
important in the discussion of the supersymmetric CP problem.

Even though the $\mu$-problem is solved perfectly, there are several
serious problems in this scenario.
Since there is no reason for the alignment of the flavor structure in
the first term in Eq.~(\ref{eq:gravity-matter}), too large rates for
flavor changing processes are predicted.
We expect flavor mixings of $O(1)$ from this form of Lagrangian.  Such
large mixings are unacceptable unless the sfermion masses are of
$O(10)$~TeV or heavier~\cite{Gabbiani:1996hi}.
The CP violating phases in the supersymmetry breaking terms are also
expected to be $O(1)$. 
In particular, a phase of the combination, $m_{1/2} \mu (B\mu)^*$, with
$m_{1/2}$ the gaugino mass and $B \mu$ defined by ${\cal L} \ni B \mu
H_u H_d + {\rm h.c.}$, cannot be eliminated by field redefinitions.  The
$\mu$ and $B \mu$ terms are generated from the multiple terms in
Eq.~(\ref{eq:gravity-higgs}) with different weights, leading to
non-aligned phases generically.
With an $O(1)$ phase for the combination, constraints from the electric
dipole moment of electron, for example, push the mass limits of
supersymmetric particles to be $O(10)$~TeV~\cite{Gabbiani:1996hi}.

There is another serious problem in cosmology. Due to the terms in
Eq.~(\ref{eq:gravity-gauge}), the scalar component of $S$ cannot carry
any (even approximately) conserving charge. In this case, there is a
moduli problem~\cite{Coughlan:1983ci,Banks:1993en}. The value of $S$
after the inflation is displaced from the minimum due to the deformation
of the $S$ potential during inflation, and at a later time $S$ finds its
true minimum and starts coherent oscillation about the true minimum. The
energy density of the oscillation then dominates over the universe
unless the displacement is much smaller than the Planck scale.
The decay of $S$, in turn, either destroys the success of the
BBN~\cite{Coughlan:1983ci} or overproduce gravitinos~\cite{Ibe:2006am}
depending on the mass range of $m_S$ (see
\cite{Dine:1983ys,Joichi:1994ce} for earlier works).
There is no range of $m_S$ which is consistent with the
cosmology~\cite{Ibe:2006am}.

It has been widely accepted that the lightest neutralino accounts for
dark matter of the universe in gravity mediation scenarios. Abundance of
thermally produced neutralinos can be calculated without information on
the detail history of the universe, and we obtain the correct order of
magnitude. However, once we take into account the existence of the
superpartner of the Goldstino, $S$, which always exists, the successful
cosmology is spoiled. We argue that an assumption made in the standard
calculation that the universe was normal up to temperatures of
$O(100)$~GeV is inconsistent with the structure of underlying models.

\subsubsection{Gauge mediation}

Some of shortcomings in gravity mediation can be cured in gauge
mediation models.
We assume in gauge mediation that the size of the supersymmetry
breaking, $F_S$ (and therefore $m_{3/2}$), is much smaller than that in
gravity mediation. The contributions to the soft supersymmetry breaking
terms come from
\begin{eqnarray}
 K_{\rm gauge}^{\rm (matter)} 
= - \frac{4 g^4 N_{\rm mess}}{(4 \pi)^4} C_2(R) ( \log | S | )^2 \Phi^\dagger \Phi\ ,
\label{eq:gauge-matter}
\end{eqnarray}
\begin{eqnarray}
 f_{\rm gauge} = 
{1 \over 2}
\left(
\frac{1}{g^2}
- \frac{2 N_{\rm mess}}{(4 \pi)^2} \log S
\right) W^\alpha W_\alpha\ ,
\label{eq:gauge-kin}
\end{eqnarray}
where $C_2(R)$ is the quadratic Casimir factors for fields $\Phi$.
These terms are generated by integrating out $N_{\rm mess}$ numbers of
messenger fields, $f$ and $\bar f$ in the fundamental representation,
which have couplings to $S$ in the superpotential, $W \ni k S f \bar
f$~\cite{Giudice:1997ni}.  Singularities at $S=0$ indicate that the
messenger fields become massless at the point.\footnote{In this
discussion, we have defined the origin of $S$ to be the point where the
messenger particles become massless. It is not necessarily the same
definition in subsection~\ref{sec:s-sector}. We will discuss a whole
set-up together with the $S$-sector shortly.} Therefore, the theory
makes sense only if the potential of $S$ has a (local) minimum at $S
\neq 0$.
Low energy parameters depend on the coupling constant $k$ only through a
logarithmic function. The dependence is encoded as the messenger scale
$M_{\rm mess} = k \langle S \rangle$ at which gauge mediation effects
appear.

The contributions from gauge mediation are much larger than those from
gravity mediation in
Eqs.~(\ref{eq:gravity-matter},\ref{eq:gravity-higgs},\ref{eq:gravity-gauge})
provided that the value of $S$ is stabilized at $S \ll M_{\rm Pl}$.
Since there is no flavor dependent terms in Eq.~(\ref{eq:gauge-matter}),
due to the flavor blindness of the gauge interactions, constraints from
flavor violating processes can be easily satisfied when $m_{3/2}
\lesssim O(1)$~GeV.

Another interesting feature is the enhancement of the $S$ couplings to
the MSSM particles~\cite{Ibe:2006rc}. The scalar component of $S$ now
has couplings to gauginos, $\lambda$:
\begin{eqnarray}
 {\cal L} \ni \frac{m_{1/2}}{\langle S \rangle} S \lambda \lambda
+ {\rm h.c.},
\end{eqnarray}
which can be much larger than the coupling to gravitinos, $\psi_{3/2}$:
\begin{eqnarray}
 {\cal L} \ni \frac{F_S^\dagger}{\Lambda^2} S^\dagger \psi_{3/2} \psi_{3/2} +
  {\rm h.c.}\ ,
\end{eqnarray}
depending on the value of $\langle S \rangle$.
Therefore, the branching fraction of the $S$ decay into gravitinos is
suppressed and the situation of gravitino overproduction from the $S$
decay can be ameliorated.

However, unfortunately, by lowering the gravitino mass $m_{3/2}$, we
have lost the natural mechanism for generating a $\mu$-term. The
contributions from Eq.~(\ref{eq:gravity-higgs}) to the $\mu$-term are
too small.
It is possible to obtain a correct size of $\mu$-term by assuming a
direct coupling between $S$ and Higgs fields such as
\begin{eqnarray}
 W \ni \epsilon S H_u H_d,
\end{eqnarray}
with a small coefficient $\epsilon$~\cite{Dine:1994vc}.
This term, however, predicts too large $B \mu$ term, $B\mu / \mu \sim (4
\pi)^2 m_{1/2}$, which is unacceptable from the electroweak symmetry
breaking. The conclusion is the same if we try to generate a $\mu$-term
from K\"ahler terms, e.g.,
\begin{eqnarray}
 K \ni \frac{1}{(4 \pi)^2} \frac{S^\dagger}{S} H_u H_d + {\rm h.c.}
\end{eqnarray}
This term can be generated by integrating out messenger fields if there
is an interaction between the Higgs and the messenger
fields~\cite{Dvali:1996cu,Giudice:1998bp}.
Although this term induces the correct size of $\mu$-term, $\mu \sim
m_{1/2}$, $B \mu / \mu$ is again larger than $m_{1/2}$ by a one-loop
factor.\footnote{There is a logical possibility of generating terms like
\begin{eqnarray}
 K \ni \frac{1}{(4 \pi)^2} H_u H_d \log |S| + {\rm h.c.}\ ,
\nonumber
\end{eqnarray}
if the Higgs fields have interactions with messenger fields.
In this case, the $B \mu$-term is not generated, whereas the $\mu$-term
is generated with the same size as the gaugino masses. This is perfectly
consistent with the electroweak symmetry breaking and also the absence
of the CP phase in $m_{1/2} \mu (B \mu)^*$ as $B \mu = 0$. This is the
only term which can be written down if $S$ carries an approximately
conserving charge. However, the authors are not aware of an explicit
model to realize this situation.}

The $\mu$-problem in gauge mediation models cannot be solved by going to
the next to minimal supersymmetric standard model (NMSSM). The correct
electroweak symmetry breaking is not achieved without further extensions
of the model~\cite{Dine:1993yw,deGouvea:1997cx}.

We cannot discuss the supersymmetric CP problem without specifying the
mechanism for $\mu$-term generation because the physical phase
arg($m_{1/2} \mu (B \mu)^*$) is not determined.

Although it is slightly model dependent, there is another issue in gauge
mediation models. In many supersymmetry breaking models, $S$ carries a
conserving charge. For example, in the O'Raifeartaigh model, there is an
unbroken $R$-symmetry where $S$ carries charge 2. In this case, as
discussed in subsection~\ref{sec:s-sector}, the $S$ field is stabilized
at $S=0$ where we cannot integrate out the messenger fields (it is the
singular point of the effective Lagrangian).\footnote{The origin of $S$
is now uniquely determined once we assign a charge to $S$.}
Additional model building efforts to shift the minimum of the $S$
potential have been needed in this type of models.

More explicitly, what we need is to spontaneously or explicitly break
the $R$-symmetry in supersymmetry breaking models. If we break it
explicitly in the Lagrangian, the theorem of~\cite{Nelson:1993nf} says
that a supersymmetric minimum appears somewhere in the field
space. Recently, there have been extensive studies on this subject, and
many simple models with explicit breaking of $R$-symmetry have been
proposed~\cite{Kitano:2006wz,Kitano:2006xg,Murayama:2006yf,Aharony:2006my}
by allowing a meta-stable supersymmetry breaking
vacuum~\cite{Intriligator:2006dd}. (See
also~\cite{Dine:2006gm,Dine:2006xt,Csaki:2006wi} for recent models with
spontaneous breaking of $R$-symmetry.)
An obvious possibility is to add an $R$-breaking cubic term in
Eq.~(\ref{eq:s-kahler}) with a small coefficient $\epsilon$,
\begin{eqnarray}
 \delta K = \frac{\epsilon}{\Lambda} S^\dagger S^2 + {\rm h.c.}
\end{eqnarray}
This term shifts the minimum of $S$ to $S = \epsilon \Lambda / 2$. This
is equivalent to give a small mass term to the messenger fields $W \ni
\epsilon \Lambda f \bar f$~\cite{Murayama:2006yf} by a field
redefinition $(S - \epsilon \Lambda/2) \to S $.
A small mass term for $S$, $W \ni \epsilon S^2$ also shifts the minimum
of $S$.

In fact, it has been known that these {\it ad hoc} deformations of the
model were not necessary once we take into account supergravity
effects~\cite{Kitano:2006wz}. The gravity mediation effects generate a
linear term of $S$ in the potential,
\begin{eqnarray}
 V \ni 2 m_{3/2} m^2 S + {\rm h.c.}
\end{eqnarray}
This is a soft supersymmetry breaking term associated with the linear
term in the superpotential in Eq.~(\ref{eq:s-linear}). By balancing with
the mass term, $V \ni m_S^2 |S|^2$ with $m_S$ in Eq.~(\ref{eq:s-mass}), we
obtain
\begin{eqnarray}
 \langle S \rangle = \frac{\sqrt 3 }{6} \frac{\Lambda^2}{M_{\rm Pl}}\ .
\label{eq:s-shift}
\end{eqnarray}
This shift is due to the fact that $R$-symmetry must be broken
explicitly in the supergravity Lagrangian (by the constant term in the
superpotential) in order to cancel the cosmological
constant~\cite{Banks:1993en}.
By taking a large $\Lambda$, the shift can be arbitrarily large.
Note here that the shift is not suppressed by the gravitino mass which
characterizes the effects of gravity mediation.  This phenomenon has
been known as the tadpole problem for singlet
fields~\cite{Hall:1983iz,Ellwanger:1983mg,Bagger:1993ji}. Small soft
supersymmetry breaking terms destabilize the hierarchy if there is a
gauge singlet field. However, this is not a problem at all for the field
$S$ and it is even better in gauge mediation to have a large enough
vacuum expectation value of $S$. Since this effect always exists, it is
the most economical way of having $S \neq 0$.

\subsubsection{Anomaly mediation}

If there is no direct coupling between $S$ and the MSSM particles even
including Planck scale suppressed operators, the leading contribution to
the sfermion/gaugino masses comes from anomaly mediation effects:
\begin{eqnarray}
 m_{1/2} = \frac{g^2 b}{(4 \pi)^2} m_{3/2}\ ,\ \ 
 m_{\rm scalar}^2 = \frac{1}{2} {d \gamma \over d \log \mu_R } m_{3/2}^2\ ,
\end{eqnarray}
where $b$ and $\gamma$ are the beta function coefficient and the
anomalous dimension, respectively, and $\mu_R$ is the renormalization
scale~\cite{Randall:1998uk}. For having $m_{1/2} = O(100)$~GeV, a large
gravitino mass $m_{3/2} \sim 10-100$~TeV is needed.
There are several good features of this scenario. Because of flavor
blindness of the mediation mechanism, there is no supersymmetric flavor
problem. The large value of $m_{3/2}$ enhances the decay rate of the
gravitino. This makes the gravitino cosmologically harmless as it decays
before the BBN starts. The cosmological moduli problem is also
absent. The $S$ field can have any conserving charges, and thus it is
reasonable to assume that $S$ has stayed at the symmetry enhanced point,
$S=0$, during and after inflation so that there is no large initial
amplitude.

Unfortunately, the minimal model is inconsistent with the
observation. The scalar leptons have tachyonic masses which would cause
a spontaneous breaking of U(1)$_{em}$ and makes the photon
massive. Therefore, we need a modification of the model.

Also there is a $\mu$-problem which is very similar to the situation in
gauge mediation. One can assume small couplings between $S$ and the
Higgs fields to give a $\mu$-term, but it causes a too large $B
\mu$-term, $B\mu / \mu \sim m_{3/2}$ with $m_{3/2} \sim 10-100$~TeV
which is unacceptable.

\subsection{Sweet Spot Supersymmetry}
\label{sec:sss}

We have encountered many problems in gravity, gauge and anomaly
mediation models. Those are summarized as follows:
\begin{itemize}
 \item Gauge mediation ($m_{3/2} \ll 100$~GeV)\\
Problems: $\mu$, (CP),
 \item Gravity mediation ($m_{3/2} \sim 100$~GeV)\\
Problems: Flavor, CP, moduli,
 \item Anomaly mediation ($m_{3/2} \sim 10-100$~TeV)\\
Problems: $\mu$, tachyonic sleptons, (CP).
\end{itemize}

There have been many attempts to circumvent these problems. For example,
in Ref.~\cite{Dine:1993yw,deGouvea:1997cx} it has been proposed to
extend a model of gauge mediation to the NMSSM by introducing a new
singlet field. However, for the successful electroweak symmetry
breaking, further extension of the model were necessary such as
introduction of vector-like matters. Similar attempts have been done in
Ref.~\cite{Chacko:1999am,Ibe:2004gh} in anomaly mediation models.
The gaugino mediation~\cite{Kaplan:1999ac} is a variance of the gravity
mediation and known to be a successful framework for solving the flavor
problem. However, since the model relies on the $SW^\alpha W_\alpha$
term for the gaugino masses, the moduli problem and the CP problem
remain unsolved.
In Ref.~\cite{Pomarol:1999ie}, a mixture of anomaly and gauge mediation
is proposed as a solution to the tachyonic slepton problem (see also
\cite{Chacko:2001jt}). The idea is to modify the structure of the
anomaly mediation by introducing an additional light degree of freedom,
$X$. It is claimed that the $\mu$-problem and the tachyonic slepton
problem can be solved by assuming appropriate couplings of $X$ to the
messenger and Higgs fields. However, it is unclear whether such a light
degree of freedom is consistent with cosmological history.

It is interesting to notice here that gauge and gravity mediation
scenarios do not share problems. This fact motivates us to think of
theories in between gauge and gravity mediation. The idea is to solve
flavor and moduli problem by reducing $m_{3/2}$, and solve the
$\mu$-problem in a similar fashion to the gravity mediation models. The
CP problem can also be solved because we can have an approximate PQ
symmetry to forbid the $B \mu$-term so that arg$(m_{1/2} \mu (B \mu)^*)
= 0$.

\begin{figure}[t]
\begin{center}
  \includegraphics[width=15cm]{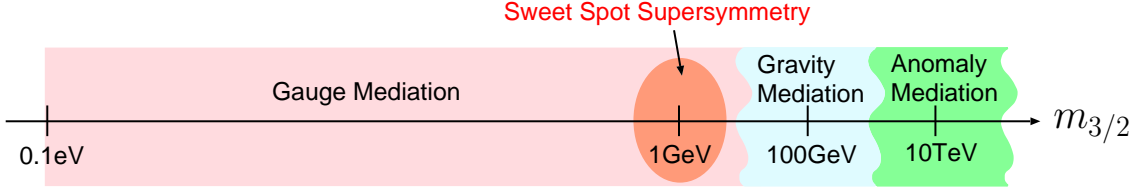}
\end{center}
\caption{Schematic picture of mediation mechanisms. Different mechanism
 works for different values of gravitino masses. A sweet spot exists at
 $m_{3/2} \sim 1$~GeV where there is no phenomenological or cosmological
 problem.} \label{fig:sweet-spot}
\end{figure}

The sweet spot exists at $m_{3/2} \sim 1$~GeV (see
Fig.~{\ref{fig:sweet-spot}}). The interaction terms between matter/gauge
field and $S$ are the same as those in gauge mediation
(Eqs.~(\ref{eq:gauge-matter},\ref{eq:gauge-kin})). For $m_{3/2} \sim
1$~GeV, possible flavor violating contributions from gravity mediation
in Eq.~(\ref{eq:gravity-matter}) are sufficiently small.
The couplings of $S$ to Higgs fields are
\begin{eqnarray}
 K_{\rm sweet}^{\rm (Higgs)}
= 
\left( \frac{S^\dagger H_u H_d}{\Lambda} + {\rm h.c.} \right)
- 
\frac{ S^\dagger S ( H_u^\dagger H_u + H_d^\dagger H_d ) }{ \Lambda^2
}\ .
\label{eq:sweet-higgs}
\end{eqnarray}
Here we replaced the Planck scale in Eq.~(\ref{eq:gravity-higgs}) with
the ``cut-off'' scale $\Lambda$ introduced in Eq.~(\ref{eq:s-kahler}).
The correct size of $\mu$-term is obtained if $\Lambda \sim
10^{16}$~GeV.
The form of the K\"ahler potential is implicitly suggesting that the
Higgs fields have some interactions with the supersymmetry breaking
sector mediated by particles with masses of $O(\Lambda)$.
We also assumed that there is an approximate PQ symmetry discussed in
subsection~\ref{sec:mssm} with $PQ(S) = 2$. With the PQ symmetry, we
cannot write down any other terms. Since $S$ carries a charge, the
K\"ahler potential for $S$ is restricted to be the form in
Eq.~(\ref{eq:s-kahler}).
The term in Eq.~(\ref{eq:s-linear}) represents
the explicit but small breaking of the PQ symmetry.

We here summarize the set-up. We consider the effective Lagrangian
written in terms of the Goldstino multiplet $S$ and the MSSM
matter/gauge fields:
\begin{eqnarray}
 K &=& S^\dagger S - \frac{c_S (S^\dagger
  S)^2}{\Lambda^2} \nonumber \\
&& + \left( \frac{ c_\mu S^\dagger H_u H_d}{\Lambda} + {\rm h.c.} \right)
- 
\frac{ c_H S^\dagger S ( H_u^\dagger H_u + H_d^\dagger H_d ) }{ \Lambda^2
} \nonumber \\
&&
+ \left(
1 - \frac{4 g^4 N_{\rm mess}}{(4 \pi)^4} C_2(R) ( \log |S| )^2 
\right)
\Phi^\dagger \Phi\ 
\ , \nonumber \label{eq:set-up}\\
&&\\
W &=& W_{\rm Yukawa} (\Phi) + m^2 S + w_0 \ , \nonumber \\
&& \nonumber \\
f &=& 
{1 \over 2}
\left(
\frac{1}{g^2}
- \frac{2 N_{\rm mess}}{(4 \pi)^2} \log S
\right) W^\alpha W_\alpha
\nonumber \ .
\end{eqnarray}
The chiral superfield $\Phi$ represents the matter and the Higgs
superfields in the MSSM, and $W_{\rm Yukawa}$ is the Yukawa interaction
terms among them.
We defined $O(1)$ valued coefficients $c_S$, $c_\mu$, and $c_H$. We
normalize the $\Lambda$ parameter so that $c_S = 1$ in the following
discussion.
The parameters $c_H$ and $\Lambda$ take real values whereas $c_\mu$ is a
complex parameter.
We consider the supergravity Lagrangian defined by the above K\"ahler
potential $K$, superpotential $W$, and gauge kinetic function $f$.
This is a closed well-defined system. The linear term of $S$ in the
superpotential represents the source term for the $F$-component of
$S$. 
The last term in the superpotential, $w_0$, is a constant, $|w_0| \simeq m^2
M_{\rm Pl}/\sqrt 3 $, which is needed to cancel the cosmological
constant.
The scalar potential has a minimum at $\langle S \rangle \sim \Lambda^2
/ M_{\rm Pl}$ which avoids the singularity at $S = 0$.
The set-up includes the dynamics of supersymmetry breaking and
mediation. By expanding fields from their vacuum expectation values, we
can obtain all the mass spectrum and interaction terms.

When we write down the Lagrangian of the standard model we usually
include the Higgs potential, $V = (\lambda_H/4)( |H|^2 - v^2 )^2$, and
the gauge interaction terms of the Higgs boson instead of just giving
bare mass terms to the $W$ and $Z$ bosons.
Analogous to that, the system above contains dynamics of the
supersymmetry breaking and a mechanism of its mediation instead of
simply writing down soft supersymmetry breaking terms.\footnote{Our
construction should not be confused with the spurion method of writing
down the soft terms. The field $S$ is a propagating field and obeys the
equation of motion.} In this sense, this way of construction is
essential for the model to be called {\it the} MSSM in a true meaning.

The effective Lagrangian is defined at the scale where the messenger
fields are integrated out. The messenger scale, $k \langle S
\rangle$, is not necessary to be $O(\langle S \rangle)$.
The $k$ parameter originally comes from superpotential terms like, $W
\ni k S f \bar f$. If the $S$ field is a composite operator above the
scale $\Lambda$ as is often the case in dynamical supersymmetry breaking
scenarios, the $k$ parameter is suppressed by a factor of
$(\Lambda/M_{\rm Pl})^{d(S)-1}$, where $d(S)$ is the dimension of the
operator $S$ above the scale $\Lambda$. Therefore, the size of $k$
depends on the actual mechanism of the supersymmetry breaking.

We can see very nontrivial consistencies in this simple set-up. First,
the $\mu$-term is generated by the K\"ahler term, $S^\dagger H_u H_d /
\Lambda$:
\begin{eqnarray}
 \mu = \frac{c_\mu F_S}{\Lambda} 
\sim m_{3/2} \left( \frac{M_{\rm  Pl}}{\Lambda} \right)\ .
\label{eq:mu}
\end{eqnarray}
With the shift of $\langle S \rangle$ in Eq.~(\ref{eq:s-shift}), the
gaugino masses are
\begin{eqnarray}
 m_{1/2} = {g^2 \over (4 \pi)^2} {F_S \over \langle S \rangle}
=
\frac{ g^2}{(4 \pi)^2} \cdot 6 m_{3/2} 
\left( \frac{M_{\rm  Pl}}{\Lambda} \right)^2\ .
\label{eq:gaugino}
\end{eqnarray}
Here and hereafter, we take a minimal model with $N_{\rm mess} = 1$. The
qualitative discussion does not change for different values of $N_{\rm mess}$.
Similar sizes of scalar masses are obtained from the K\"ahler terms.
Finally, the moduli problem now turns into a mechanism for the
production of dark matter. The energy density of the coherent
oscillation of $S$ dominates over the universe, and the reheating
process by decays of the $S$-condensation later produces gravitinos
through a rare decay process $S \to \psi_{3/2} \psi_{3/2}$. The amount
can be expressed in terms of $m_{3/2}$ and $\Lambda$~\cite{Ibe:2006rc}:
\begin{eqnarray}
 \Omega_{3/2} h^2 = 0.1
\times 
\left(\frac{m_{3/2}}{500~{\rm MeV}}\right)^{3/2}
\left(\frac{\Lambda}{1 \times 10^{16}~{\rm GeV}}\right)^{3/2}\ .
\label{eq:omega}
\end{eqnarray}
Here we have assumed that the decay of $S$ into two Higgs bosons, $S \to
h h$, is the dominant decay channel.
The phenomenological requirements that $\mu \sim m_{1/2} \sim
O(100)$~GeV, and $\Omega_{3/2} h^2 \simeq 0.1$ can all be satisfied when
$m_{3/2} \sim 1$~GeV and $\Lambda \sim 10^{16}$~GeV.

\begin{figure}[t]
\begin{center}
  \includegraphics[height=8.5cm]{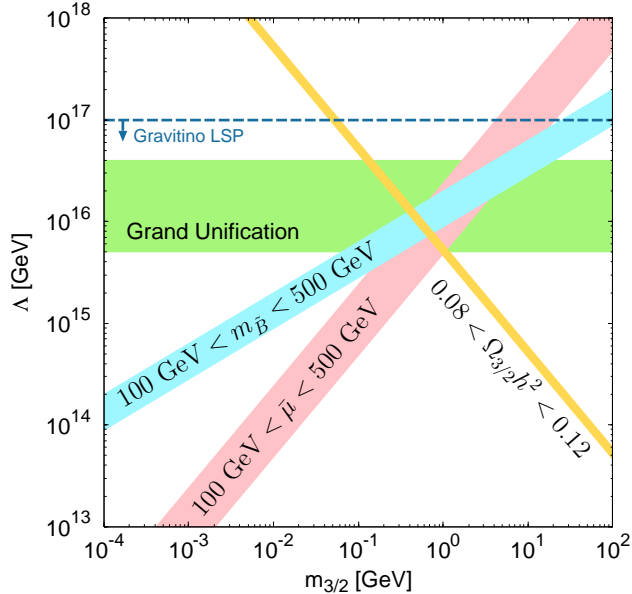}
\end{center}
\caption{Phenomenologically required values of the Higgsino mass $\bar
 \mu$ (with an $O(1)$ ambiguity, see text), the Bino mass $m_{\tilde B}$
 and the gravitino energy density $\Omega_{3/2} h^2$. These three
 quantities have different dependencies on parameters $m_{3/2}$ and
 $\Lambda$. The three bands meet around $m_{3/2} \sim 1$~GeV and
 $\Lambda \sim M_{\rm GUT}$. The quantity $\Omega_{3/2} h^2$ is defined
 in Eq.~(\ref{eq:omega}). It represents the energy density of the
 non-thermally produced gravitinos through the decays of $S$ if $S \to
 hh$ is the dominant decay channel.}
\label{fig:unif}
\end{figure}

We can see the non-trivial success of this framework in
Fig.~\ref{fig:unif}, where we see how $O(1)$~GeV gravitino mass is
selected. The bands of $100~{\rm GeV} < \bar \mu < 500~{\rm GeV}$,
$100~{\rm GeV} < m_{\tilde B} < 500~{\rm GeV}$, and $0.08 < \Omega_{3/2}
h^2 < 0.12$ are shown, where we defined $\bar \mu \equiv m_{3/2} M_{\rm
Pl} / \Lambda$ and $\Omega_{3/2} h^2$ by
Eq.~(\ref{eq:omega}).\footnote{The band of $\Omega_{3/2} h^2$ does not
represent the dark matter density once we deviate far from the region of
$m_{3/2} \sim 1$~GeV. For $m_{\tilde B} \ll \bar \mu$ or $m_{\tilde B}
\gg \bar \mu$, the successful electroweak symmetry breaking cannot be
achieved, and we cannot perform a sensible calculation of the $S \to h
h$ decay width.}
The Bino mass $m_{\tilde B}$ is the mass of the U(1)$_Y$ gaugino.
Surprisingly, these {\it three} bands meet at $m_{3/2} \sim 1$~GeV and
$\Lambda \sim M_{\rm GUT} \sim 10^{16}$~GeV.

The fact that $\Lambda$ coincides with the unification scale, $M_{\rm
GUT}$, is also quite interesting. In grand unified theories (GUTs), such
as in SU(5) or SO(10) models, we need to introduce colored Higgs fields
in order for models to be consistent with gauge invariance. The colored
Higgs fields, however, need to get masses through the spontaneous
breaking of SU(5) or SO(10). This suggests that the Higgs multiplets
have some interactions with the GUT-breaking sector whose typical mass
scale is, of course, $M_{\rm GUT}$. Therefore, it is quite natural to
have $M_{\rm GUT}$ suppressed interactions in the low energy effective
theory. The same ``cut-off'' scale $\Lambda$ for $S$ suggests that the
dynamics of GUT breaking is responsible for the supersymmetry breaking
as well. The picture of unification of the Higgs sector, the
supersymmetry breaking sector and the GUT breaking sector naturally
comes out.
Although it sounds like a very ambitious attempt to build a realistic
model to realize this situation, it is quite possible and even very
simple to build such a dream model by using a recent theoretical
development of supersymmetric field
theories~\cite{Intriligator:2006dd}. For an explicit example of such a
GUT model, see Ref.~\cite{Kitano:2006wm}.

This quite simple framework summarized in Eq.~(\ref{eq:set-up}), gauge
mediation with direct couplings between supersymmetry breaking sector
$S$ and the Higgs fields at the GUT scale, solves all the problems we
mentioned before. We discuss these one by one here.

\subsubsection*{Supersymmetric flavor problem}

       The gravity mediated contributions to the sfermion masses squared
       are of $O(m_{3/2}^2)$. Therefore, the flavor mixing in sfermions
       are at most of $O(10^{-4})$ level for $m_{3/2} \sim 1$~GeV. For
       example, the constraints from the $\mu \to e \gamma$ decay and
       $\mu \to e$ conversion process in nuclei put bounds on the mixing
       to be~\cite{Gabbiani:1996hi}
\begin{eqnarray}
 (\delta^l_{12})^{\rm eff}_{LR, RL} 
\sim  
\left( 
\frac{m_{3/2}^2}{m_{\rm SUSY}^2}
\right)
\left(
\frac{m_\mu \tan \beta}{m_{\rm SUSY}}
\right)
\lesssim 10^{-6}\ ,
\end{eqnarray}
where $m_{\rm SUSY}$ is a typical sfermion/gaugino mass scale and
$m_\mu$ is the muon mass.
       By using the fact that the value of $\tan \beta\ (\equiv \langle
       H_u \rangle / \langle H_d \rangle)$ is predicted to be $O(30-40)$
       (see discussion in the next section), this bound is marginally
       satisfied with sfermion masses of $O(100)$~GeV.
       If gravitational dynamics maximally violates flavor conservation,
       future or on-going experiments have good chances to see the
       effects~\cite{Ritt:2006cg,Kuno:2005mm}.

       The flavor mixings from the high-scale dynamics such as physics
       at the GUT scale~\cite{Hall:1985dx} and the effect of
       right-handed neutrinos~\cite{Borzumati:1986qx} are small as is
       always the case in gauge mediation.

\subsubsection*{Supersymmetric CP problem}

       There are two physical phases in the MSSM:
\begin{eqnarray}
 \arg (m_{1/2} \mu (B \mu)^*)\ ,\ \ \ \arg (m_{1/2} A^*)\ .
\end{eqnarray}
       From the K\"ahler term in Eq.~(\ref{eq:sweet-higgs}), $A$- and
       $B$-terms of $O(m_{3/2})$ are generated, but these will be
       overwhelmed by one-loop renormalization group (RG) contributions
       below the messenger scale. Since the RG contributions are
       proportional to the gaugino masses, the physical phases above are
       approximately vanishing.

       In fact, the phases of the original $O(m_{3/2})$ contributions
       are also aligned with those of gaugino masses. The phases of
       three complex parameters in the Lagrangian, i.e., $m^2$, $c_\mu$,
       and $w_0$, can all be taken to be the same by a field
       redefinition via U(1)$_R$ and U(1)$_{PQ}$ transformations.

       Even if there are $O(1)$ phases in the $O(m_{3/2})$
       contributions, which is possible if the PQ symmetry is maximally
       violated by operators suppressed by the Planck scale, the above
       physical phases are of $O(1\%)$ which again marginally satisfies
       the experimental constraints. The upper bound on the electric
       dipole moment of the electron, for example, gives a
       constraint~\cite{Gabbiani:1996hi}:
\begin{eqnarray}
 \left(
m_{3/2} \over m_{\rm SUSY}
\right)
\left(
m_e \tan \beta \over m_{\rm SUSY}
\right)
 \lesssim 10^{-7}\ ,
\end{eqnarray}
where $m_e$ is the electron mass.  The bound corresponds to $m_{\rm
       SUSY} \gtrsim 300$~GeV for $m_{3/2} \sim 1$~GeV.

\subsubsection*{{\boldmath $\mu$}-problem}

       There are three kinds of $\mu$-problem in the MSSM, i.e., ``~Why
       $\mu \ll M_{\rm Pl}$?'', ``~Why $\mu^2 \sim m_{H_u}^2$?'', and
       ``~Why $\mu \sim m_{1/2}$?'' The second and third ones are
       related because there is a one-loop correction to the $m_{H_u}^2$
       parameter proportional to $m_{1/2}^2$.

       The first one was answered by the approximate PQ symmetry. The
       $\mu$-term is forbidden by symmetry, but induced by a small
       explicit breaking term, $W \ni m^2 S$.

       We can naturally obtain the relation, $\mu^2 \sim m_{H_u}^2$,
       once we assume the form of the K\"ahler potential to be the one
       in Eq.~(\ref{eq:sweet-higgs}). This is a generalization of the
       Giudice-Masiero mechanism in gravity
       mediation~\cite{Giudice:1988yz}. The relation is independent of
       the ``cut-off'' scale $\Lambda$. We discuss a possible origin of
       the K\"ahler terms later.

       The final relation, $\mu \sim m_{1/2}$, is realized when $\Lambda
       \sim M_{\rm GUT}$ as we can see in Fig.~\ref{fig:unif}. From
       Eq.~(\ref{eq:mu}) and (\ref{eq:gaugino}), the relation between
       $\bar \mu$ and the Bino mass, $m_{\tilde B}$, is
\begin{eqnarray}
 {\bar \mu \over m_{\tilde B}} = 0.6 \times
\left(
{\Lambda \over {1 \times 10^{16}~{\rm GeV}}}
\right)\ .
\end{eqnarray}
       Although it is an `accident' to have similar values of $\mu$ and
       the gaugino masses, the value we need, $\Lambda \sim M_{\rm
       GUT}$, is motivated by two other independent physics, i.e., grand
       unification and dark matter of the universe.

\subsubsection*{Cosmological moduli/gravitino problem}

       As we have already discussed, the energy density carried by the
       coherent oscillation of $S$ would not cause a problem. The decay
       of $S$ reheats the temperature of the universe to of $O(100)$~MeV
       for $m_{3/2} \sim 1$~GeV and $\Lambda \sim 10^{16}$~GeV. This is
       high enough for the standard BBN. The non-thermal gravitino
       production from this decay gives the largest contribution to the
       matter energy density of the universe. The amount in
       Eq.~(\ref{eq:omega}) is, amazingly, consistent with the
       observation.

       The baryon asymmetry existed before $S$ decays is diluted by the
       entropy production. If we assume the initial amplitude of $S$ to
       be of $O(\Lambda)$, the dilution factor is estimated to be of
       order $10^{-4} (T_R/10^8~{\rm GeV})^{-1}$ with $T_R$ the
       reheating temperature after inflation. Therefore, a larger amount
       of baryon asymmetry is needed to be generated if baryogenesis
       happened above the temperature of $O(100)$~MeV.

       If the stau is the NLSP as in the case we will study later, staus
       are also non-thermally produced through the $S$ decays if it is
       kinematically allowed. The pair annihilation process reduces the
       amount but the abundance ends up with of $O(50)$ times larger
       than the result of the standard calculation of the thermal relic
       abundance. There are constraints on the decay of the staus into
       gravitinos from the BBN. Recent calculations including the
       catalyzing effects give an upper bound on the life-time of stau
       to be
       $O(1000)$~seconds~\cite{Pospelov:2006sc,Kohri:2006cn,Cyburt:2006uv,Hamaguchi:2007mp}. Although
       the lifetime is extremely sensitive to the stau mass $(\propto
       m_{\tilde \tau}^5)$, the typical lifetime with $m_{3/2} \sim
       1$~GeV is on the border of this constraint. This coincidence may
       be interesting for the Lithium abundance of the
       universe~\cite{Kohri:2006cn,Kaplinghat:2006qr}.

\subsubsection*{Unwanted axion?}

It is common in gauge mediation scenarios that there is an approximate
U(1) symmetry which is spontaneously broken. Therefore there is a
(possibly unwanted) Goldstone boson associated with
it~\cite{Banks:1993en}. In the scenario we are discussing the vacuum
expectation value of $\langle S \rangle$ breaks the approximate PQ
symmetry spontaneously.
The axion associated with the symmetry breaking is actually the scalar
component of $S$ itself. The $S$ scalar has a mass of the order of
100~GeV (see Eq.~(\ref{eq:s-mass})) because of the linear term in the
superpotential.
Interactions between PQ currents and the axion $S$ are suppressed by the
scale of the symmetry breaking $\langle S \rangle \sim
10^{14}$~GeV. There is no experimental or astrophysical constraint on
such a particle. As we discussed above, the $S$ scalar even plays an
essential role in cosmology.

\subsubsection*{Dimension-four and five proton decay problem}

       The dimension-four operators which violate the baryon number
       conservation are forbidden by an unbroken $Z_2$ subgroup of the
       PQ symmetry. This is identical to the $R$-parity.

       Dimension five operators, such as $QQQL$, are allowed to appear
       at low energy because the PQ symmetry is spontaneously broken. In
       particular, if there are following terms in the superpotential:
\begin{eqnarray}
 S QQQL\ ,\ \ \ S UUDE\ ,
\end{eqnarray}
       the dangerous terms like $QQQL$ and $UUDE$ appear by substituting
       the vacuum expectation value of $S \sim \Lambda^2/M_{\rm Pl} \sim
       10^{14}$~GeV. In GUT models, these effective operators can be
       generated by diagrams with colored-Higgs exchange.
       In this case, the coefficients of the above operators will
       typically be of $O(f_u f_d/M_{\rm GUT}^2)$ where $f_u$ and $f_d$
       are the Yukawa coupling constants of up- and down-type quarks.
       By substituting $\langle S \rangle$, this becomes effectively
       $QQQL$ or $UUDE$ operators suppressed by $f_u f_d / M_{\rm Pl}$.
       The prediction to the proton life-time is on the border of the
       experimental constraints with such
       coefficients~\cite{Goto:1998qg}. 

\subsubsection*{UV completion}

       The discussion so far is based on the low energy effective theory
       defined in Eq.~(\ref{eq:set-up}).  This effective theory is valid
       up to the messenger scale $k \langle S \rangle$. Although it is
       not necessary for the discussion of low energy physics to specify
       UV models, an existence proof of an explicit UV completion
       supports our ansatz in Eq.~(\ref{eq:set-up}).

       It is straightforward to UV complete the theory above the
       messenger scale by simply assuming a presence of messenger
       particles $f$ and $\bar f$ which carry the standard model quantum
       numbers, and an interaction term $k S f \bar f$.
       The full model is $K \ni f^\dagger f + \bar f^\dagger \bar f$ and
       $W \ni  k S f \bar f$ instead of terms involving $\log S$
       in Eq.~(\ref{eq:set-up}).

       The model with messenger fields now has a supersymmetric and
       hence stable vacuum at $S = 0$ and $f = \bar f = \sqrt{ - m^2 /
       k}$. However, as it has been shown in Ref.~\cite{Kitano:2006wz},
       there is a meta-stable minimum at $\langle S \rangle \sim
       \Lambda^2 / M_{\rm Pl}$ where supersymmetry is broken and
       messenger fields are massive. The effective theory in
       Eq.~(\ref{eq:set-up}) correctly describes physics around the
       meta-stable vacuum.

\begin{figure}[t]
\begin{center}
  \includegraphics[width=13.cm]{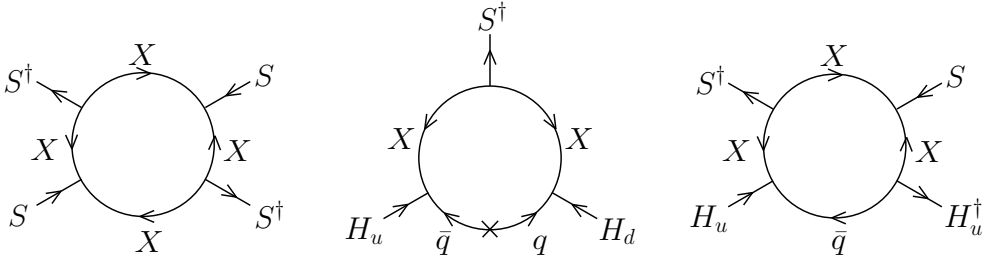}
\end{center}
\caption{Feynman diagrams to generate higher dimensional operators in a
UV model.}
\label{fig:loop}
\end{figure}

       Above the mass scale $\Lambda$, we need a further UV
       completion. The simplest model is the O'Raifeartaigh
       model~\cite{O'Raifeartaigh:1975pr}:
\begin{eqnarray}
 K = S^\dagger S + X^\dagger X + Y^\dagger Y\ ,
\end{eqnarray}
       and
\begin{eqnarray}
 W_S = m^2 S + {\kappa \over 2} S X^2 + M_{XY} X Y\ ,
\label{eq:o'rai}
\end{eqnarray}
       where $\kappa$ and $M_{XY} (\gg m)$ are a coupling constant and a mass
       for $X$ and $Y$, respectively. 
       There is an approximate PQ symmetry with charges $PQ(X)=-1$ and
       $PQ(Y)=1$.
       By integrating out massive fields $X$ and $Y$, we obtain the
       K\"ahler term $-(S^\dagger S)^2 / \Lambda^2$ with
\begin{eqnarray}
{1 \over \Lambda^{2}} = \frac{|\kappa|^4}{12 (4 \pi)^2}
 \frac{1}{M_{XY}^2}\ ,
\label{eq:lam-s}
\end{eqnarray}
       at one-loop level (see Fig.~{\ref{fig:loop}}).  
       The Higgs fields can directly couple to this system so that we
       obtain effective operators in Eq.~(\ref{eq:set-up}).
       The terms are generated by introducing following interaction
       terms in the superpotential:
\begin{eqnarray}
 W_{\rm Higgs} = h H_u \bar q X + \bar h H_d q X + M_q q \bar q\ ,
\label{eq:o'rai-higgs}
\end{eqnarray}
       where $h$ and $\bar h$ are coupling constants.  Again, the PQ
       symmetry is preserved for $PQ(q)=PQ(\bar q)=0$. The supersymmetry
       breaking still happens in this extended model.  After integrating
       out $q$ and $\bar q$, we obtain the $c_\mu S^\dagger H_u H_d /
       \Lambda$ term with
\begin{eqnarray}
 {c_\mu \over \Lambda} = - {\kappa^* h \bar h \over (4 \pi)^2}
  \frac{1}{M_q} \cdot f\left({M_{XY}^2 \over M_q^2 }\right)\ ,
\end{eqnarray}
       where 
\begin{eqnarray}
 f(x) = {1-x+\log x \over {(1-x)^2}}\ .
\end{eqnarray}
       The term $- c_H S^\dagger S H_u^\dagger H_u /
       \Lambda^2$ is also generated with
\begin{eqnarray}
 {c_H \over \Lambda^2} = {| \kappa |^2 | h |^2 \over {(4 \pi)^2}}
{1 \over M_q^2} \cdot g\left({M_{XY}^2 \over M_q^2 }\right)\ ,
\end{eqnarray}
       where
\begin{eqnarray}
 g(x) = {- 3 + 4x - x^2 - 2 \log x \over {2 (1-x)^3}}\ .
\label{eq:g-func}
\end{eqnarray}
       These are obtained by calculating Feynman diagrams in
       Fig.~\ref{fig:loop}.  No other unwanted terms are generated
       because of the approximate PQ symmetry in the model.

We can obtain the relation $c_\mu \sim c_H \sim 1$ for $M_{XY} \sim M_q$
if the values of $\kappa$, $h$ and $\bar h$ are relatively large.
       In particular, we find
\begin{eqnarray}
 {| \mu |^2 \over m_{H_u}^2}
= {| c_\mu |^2 \over c_H^2}
= { | \bar h |^2 \over ( 4 \pi )^2}
  {f(x)^2 \over g(x)}\ ,
\label{eq:ratio}
\end{eqnarray}
       where $x = M_{XY}^2 / M_q^2$.  The $\mu$-term squared is suppressed by
       a one-loop factor compared to the soft mass term $m_{H_u}^2$ for
       $M_{XY} \sim M_q$.
 The function $f(x)^2/g(x)$ never exceeds $O(1)$ values even for general
 relations between $M_{XY}$ and $M_q$.
       To avoid a too large hierarchy, the loop expansion parameter $|
       \bar h |^2 / (4 \pi)^2$ should not be too small, i.e., the model
       should be (semi) strongly coupled.\footnote{Note that this is not
       the same situation as the discussion around Eq.~(20). The
       one-loop factor enhancement there, $B \mu / \mu = m_{1/2} / (g^2/
       (4 \pi)^2)$, is always large due to the perturbativity of the
       standard model gauge coupling $g$.}
       This fact suggests that this O'Raifeartaigh model itself is an
       effective theory of some dynamical supersymmetry breaking models.

Indeed, there is an incredibly simple dynamical model which provides the
above O'Raifeartaigh model as an effective description. The model is
also embeddable into an SU(5) unified model in a straightforward
way. The same dynamics spontaneously breaks SU(5) gauge symmetry and
supersymmetry~\cite{Kitano:2006wm}.

\begin{figure}[t]
\begin{center}
  \includegraphics[width=13.cm]{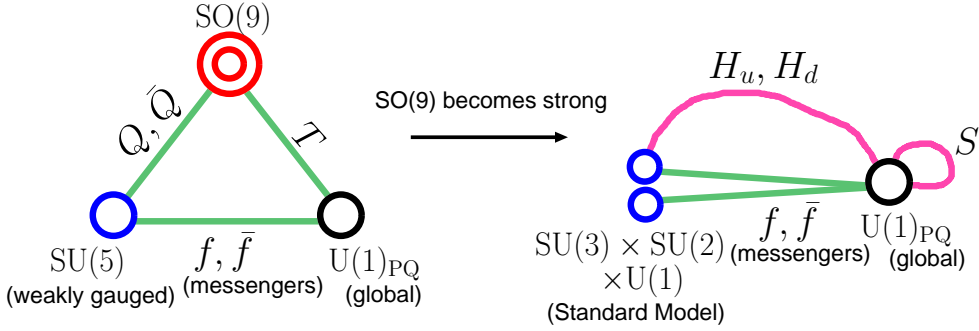}
\end{center}
\caption{Structure of an example of the UV model~\cite{Kitano:2006wm}.}
\label{fig:quiver}
\end{figure}

       The model is based on a strongly coupled gauge theory where $S$
       and the Higgs fields appears at low energy as massless
       hadrons. The constituent `quarks' of these hadrons are $Q$, $\bar
       Q$ and $T$, all of which transform as a vector representation of
       a strong SO(9) gauge group. Also $Q$ and $\bar Q$ carry standard
       model quantum numbers (${\bf 5}$ and ${\bf \bar 5}$ under SU(5))
       and $T$ is singlet under SU(5) but carries $PQ(T) = 1$. (See
       Fig.~\ref{fig:quiver} for the structure of the model.)
       The Higgs fields and $S$ are identified with meson fields
\begin{eqnarray}
 H \sim (Q T)\ , \ \ \ \bar H \sim (\bar Q T)\ ,\ \ \ S \sim (T T)\ ,
\end{eqnarray}
       where the Higgs fields in the ${\bf 5}$ and ${\bf \bar 5}$
       representation, $H$ and $\bar H$, contains $H_u$ and $H_d$ as
       SU(2) doublet components, respectively.

\begin{figure}[t]
\begin{center}
  \includegraphics[width=14.cm]{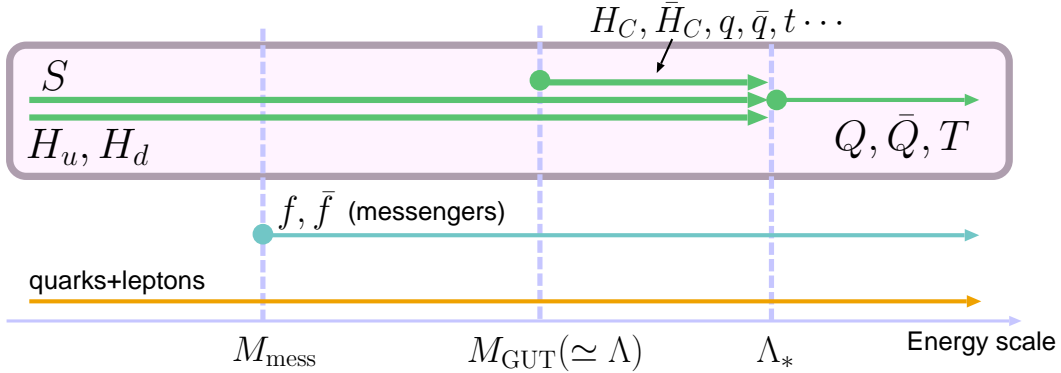}
\end{center}
\caption{Particles to describe the theory in each energy interval.}
\label{fig:description}
\end{figure}

       This is an SO(9) gauge theory with eleven flavors, and the SU(5)
       gauge group is identified with a subgroup of the SU(11) flavor
       symmetry.
       We can write down superpotential terms:
\begin{eqnarray}
 W_{\rm GUT} = \mu_T T^2 + M_Q Q \bar Q - {1 \over M_X} (Q \bar Q )^2
+ \cdots\ ,
\end{eqnarray}
       where $\mu_T$ ($\sim 1-10$~GeV) corresponds to the small explicit
       breaking of the PQ symmetry. This term is going to be the $m^2 S$
       term in Eq.~(\ref{eq:set-up}) at low energy.
       Once we ignore the superpotential (in the limit of $\mu_T, M_Q
       \to 0$ and $M_X \to \infty$), the SO(9) 11 flavor theory is on
       the edge of the conformal
       window~\cite{Intriligator:1995id}. Therefore, at some scale
       $\Lambda_*$ the gauge coupling constant flows into the infrared
       fixed point. Although it becomes a strongly coupled conformal
       field theory (CFT) near the fixed point, there is a dual weakly
       coupled CFT description with which we can perform perturbative
       calculations.
The dual gauge group is SO(6) and the superpotential above is replaced
with
\begin{eqnarray}
 W_{\rm GUT}^{\rm dual} &=& \mu_T \Lambda_* S
+ M_Q \Lambda_* M
- {\Lambda_*^2 \over M_X} M^2 + \cdots
\nonumber \\
&&
+ {\kappa_* \over 2 } S t t
+ h_* H \bar q t
+ h_* \bar H q t
+ h_* M q \bar q
+ \cdots
\label{eq:gut}
\end{eqnarray}
       The field $M$ is a composite meson $M \sim (Q \bar Q)$ which
       transforms as ${\bf 1} + {\bf 24}$ representation under SU(5).
       The fields $t$, $q$, and $\bar q$ are dual quarks which are
       charged under SO(6), and $\kappa_*$ and $h_*$ are the coupling
       constants at the fixed point ($\kappa_* = h_* \sim (4 \pi)/ N$
       with $N=6$).
       At a vacuum where the gauge group is broken down to the standard
       model gauge group, $\langle M \rangle = {\rm diag.}  (0,0,0,v,v)$
       and $\langle q_C \rangle \neq 0$ ($q_C:$ colored components of
       $q$), this model becomes exactly the same as the O'Raifeartaigh
       model in Eqs.~(\ref{eq:o'rai}) and (\ref{eq:o'rai-higgs}) with
       the identification of $t \to X$, $H_C \to Y$ ($H_C:$ the colored
       Higgs field), $\mu_T \Lambda_* \to m^2$, $h_* \langle q_C \rangle
       \to M_{XY}$ and $h_* \langle M \rangle \to M_q$. (See
       Fig.~\ref{fig:description} for particles to describe the
       effective theory in each energy interval.)

       Once we take into account non-perturbative effects, there appears
       a supersymmetric minimum far away from the origin of
       $S$. However, it has been shown in
       Ref.~\cite{Intriligator:2006dd} that the vacuum near $S=0$ is
       meta-stable. We can see in Eq.~(\ref{eq:lam-s}) that the $S$ mass
       squared, $m_S^2 = + 4 F_S^2 / \Lambda^2$, is indeed positive.

       The loop expansion parameter, $|h_*|^2 N / (4 \pi)^2$ where
       $N=6$, is $1/N_F$ at the fixed point in this
       model ($N_F = 11$). Therefore, we obtain
\begin{eqnarray}
 {\mu^2 \over m_{H_u}^2 } \sim {1 \over N_F} \ \ \ (N_F=11)\ .
\label{eq:ratio-strong}
\end{eqnarray}
       Although it looked problematic to have a hierarchy in
       Eq.~(\ref{eq:ratio}) in perturbative models, similar sizes of
       $\mu$ and $m_{H_u}$ can be obtained in this semi strongly coupled
       theory: $\mu / m_{H_u} \sim 1/3$.\footnote{The relation is not a
       precise prediction of the model. Depending on the ratio of the
       mass parameters $M_q (\equiv h_* \langle M \rangle)$ and $M_{XY}
       (\equiv h_* \langle q_C \rangle)$, which are independent
       parameters in the superpotential, there can be $O(1)$ deviation
       from the relation (see Eq.~(\ref{eq:ratio})). If $M_q \lesssim
       M_{XY}$, we can reliably use Eq.~(\ref{eq:ratio}) (by multiplying
       a factor of $N$) with $h_*^2 N/(4 \pi)^2 = 1/ N_F$ as the leading
       order result of the $1/N$ expansion. However, once $M_q / M_{XY}$
       becomes too large, such as a factor of three or so, we lose a
       perturbative control of the calculation. In this case, we can
       first integrate out $q_D$ and $\bar q_D$ (the doublet part of $q$
       and $\bar q$), and then by taking the Seiberg
       duality~\cite{Seiberg:1994pq} of this dual picture again the
       theory becomes (semi) weakly coupled. (It is an SO(5) 7 flavor
       model. See Ref.~\cite{Kitano:2006wm}.)  Although there are $O(1)$
       ambiguities in model parameters through the matching between two
       theories, the naive dimensional analysis~\cite{Luty:1997fk} gives
       the same result as the relation in Eq.~(\ref{eq:ratio-strong})
       even in that case.  }

\subsubsection*{Doublet-Triplet splitting problem}

       The model above completely solves the doublet-triplet splitting
       problem in GUT models. By the vacuum expectation value of the
       colored component of $q$ and $\bar q$, the gauge group SO(6)
       $\times$ SU(5) is broken down to the standard model gauge group.
       The $H \bar q t$ and $\bar H q t$ couplings in Eq.~(\ref{eq:gut})
       then give mass terms only for the colored Higgs
       fields~\cite{Kitano:2006wm}. This dual picture is similar to an
       SO(10) model proposed in Ref.~\cite{Hotta:1996qb}.

As discussed before, the dimension five operators for proton decays are
sufficiently suppressed thanks to the approximate PQ symmetry.

\subsubsection*{Supersymmetric fine-tuning problem}

       The experimental lower limit on the Higgs boson mass from LEP-II
       experiment, $m_h > 114$~GeV~\cite{Barate:2003sz}, has put a
       threat on supersymmetric models.
       In order to satisfy the experimental bound, we need either a
       heavy scalar top quark (stop) or a large $A_t$-term (the
       stop-stop-Higgs coupling) since a significant one-loop
       contribution to $m_h$ is necessary~\cite{Okada:1990vk}.
       On the other hand, once we have large $m_{\tilde t}$ or $A_t$, it
       induces a large one-loop contribution to the soft mass term
       $m_{H_u}^2$. This immediately means that there is a fine-tuning
       in the electroweak symmetry breaking as we can see in the
       condition:
\begin{eqnarray}
 {M_Z^2 \over 2} \simeq - \mu^2 - m_{H_u}^2 (\Lambda) 
- \delta m_{H_u}^2\ ,
\label{eq:ewsb}
\end{eqnarray}
       where $\delta m_{H_u}^2$ is the contribution from the radiative
       correction, and $M_Z$ is the $Z$ boson mass ($M_Z =
       91.2$~GeV). If $|\delta m_{H_u}^2| \gg M_Z^2$, we need
       cancellation between $\delta m_{H_u}^2$ and either $\mu^2$ or
       $m_{H_u}^2 (\Lambda)$ to reproduce a correct value of the $Z$
       boson mass. A cancellation of at least $O(1-5 \%)$ is necessary
       to satisfy the bound on the Higgs boson mass in generic gravity
       or gauge mediation models (see for review \cite{Kitano:2006gv}).

       Although the framework in Eq.~(\ref{eq:set-up}) does not avoid
       the problem, there is an interesting consistency.  As we have
       observed in an example of the UV completion before, the ratio of
       $\mu^2 / m_{H_u}^2 (\Lambda)$ is predicted to be small (at least
       a factor of a few) if the theory has (semi) perturbative
       description above the scale $\Lambda$. In general, without
       specifying UV models there is a good reason to believe that the
       description should be (semi) perturbative.
First, we are implicitly assuming that quarks and leptons remain to be
elementary particles (weakly coupled) all the way up to the Planck
scale, otherwise we reintroduce the flavor problem. If quarks and
leptons are strongly coupled above the scale $\Lambda$, it is expected
to have interaction terms such as $S^\dagger S \Phi^\dagger \Phi /
\Lambda^2$ which induce mixing terms of $O(1)$ in sfermion mass
matrices.
On the other hand, we must write down the Yukawa coupling constant for
the top quark which is of $O(1)$. This indicates that the Higgs fields
should not be replaced by a composite operator with a large
dimension. If the dimension of the Higgs operator above the scale
$\Lambda$ was $d(H) > 1$ and the top quark is an elementary particle as
discussed above, the Yukawa coupling would be suppressed by a factor of
$(\Lambda/M_{\rm Pl})^{d(H)-1}$. The $O(1)$ Yukawa coupling constant
suggests that dimension of the Higgs operator is $d(H) \simeq 1$, i.e.,
the Higgs fields are not (very) strongly coupled.
       The loop expansion should then make sense in the calculation of
       $\mu$ and $m_{H_u}^2 (\Lambda)$. The smallness of $\mu^2 /
       m_{H_u}^2 (\Lambda)$ is, therefore, a generic feature of the
       model.

 Another amusing point to notice is that the function $g(x)$ in
Eq.~(\ref{eq:g-func}) is positive valued for $x > 0$. This means
$m_{H_u}^2 (\Lambda) > 0$ if there is a (semi) perturbative description.

       Now with positive $m_{H_u}^2 (\Lambda)$ and $\mu^2 \ll m_{H_u}^2
       (\Lambda)$, the condition of the electroweak symmetry breaking in
       Eq.~(\ref{eq:ewsb}) implies that $\delta m_{H_u}^2$ must be
       negative and large. Indeed, the contributions from the stop-loop
       diagrams are negative and are proportional to $m_{\tilde
       t}^2$. Therefore, the ``little hierarchy'', i.e., a heavy stop is
       predicted in this model. The tight experimental bound on the
       Higgs boson mass is not a big surprise.

\subsubsection*{Strong CP problem}

       Although the approximate PQ symmetry introduced in the framework
       does not provide us with a solution to the strong CP problem in a
       usual way by the axion mechanism~\cite{Peccei:1977hh} (because it
       is explicitly broken), there is an interesting connection.

       The PQ symmetry is anomalous with respect to the SU(3) strong
       interaction of the standard model. If we demand the PQ symmetry
       to be non-anomalous, there are two options to take. The first one
       is to simply assume that the $u$-quark is massless. By combining
       U(1)$_{PQ}$ with the chiral symmetry, we can make the PQ symmetry
       non-anomalous.

       Another option is to introduce an axion chiral superfield $A$
       which has a coupling to the gauge fields:
\begin{eqnarray}
 f \ni {A \over f_A} W^\alpha W_\alpha\ .
\end{eqnarray}
       The kinetic term for $A$ is
\begin{eqnarray}
 K \ni (A + A^\dagger)^2\ .
\end{eqnarray}
       With a PQ transformation of $A$, $A \to A + i \theta$, we can
       cancel the gauge anomaly.

\setcounter{footnote}{0}

Each of two options, massless $u$-quark and the axion, solves the strong
CP problem.\footnote{The scalar component of the axion chiral
superfield $A$, the saxion, obtains a mass of the order of 1~GeV by gravity
mediation effects~\cite{Goto:1991gq}. With a sufficiently small decay
constant $f_A$, the saxion does not cause a moduli problem as it decays
into a pair of gluon much earlier than the decay of $S$.}

\section{Low energy predictions}
\label{sec:low-energy}

\setcounter{footnote}{0}

The set-up in Eq.~(\ref{eq:set-up}) provides a characteristic spectrum
of the supersymmetric particles. It is different from conventional gauge
or gravity mediation models. Since the Higgs sector directly couples to
the supersymmetry breaking sector at the GUT scale, the soft mass terms
for the Higgs fields are generated at the GUT scale. 
The gaugino masses and sfermion masses are, on the other hand, generated
at the messenger scale. This hybrid feature provides interesting
predictions on the low energy spectrum.

We discuss a parametrization of the model defined in
Eq.~(\ref{eq:set-up}), with which we can calculate the low energy
spectrum and interaction terms. As we will see below, we can parametrize
the model by three quantities. These three define a theoretically
well-motivated hypersurface in the large dimensional MSSM parameter
space.

\subsection{Parametrization}

We first count the number of the parameters in the model. The soft
supersymmetry breaking terms for the Higgs sector:
\begin{eqnarray}
 m_H^2\ , \ \ \ \mu\ ,
\end{eqnarray}
are generated at the scale $\Lambda$. We take the scale $\Lambda$ to be
the unification scale $M_{\rm GUT}$.
We assumed the same soft mass
terms for $H_u$ and $H_d$ ($m_{H_u}^2 (M_{\rm GUT}) = m_{H_d}^2 (M_{\rm
GUT}) = m_H^2$) as motivated by the UV completion discussed
before. Gaugino masses, $A$-terms, $B$-term, and sfermion masses are
vanishing at the GUT scale.

Below the GUT scale, RG evolutions of the soft terms induce sfermion
masses through the Yukawa interactions. The gaugino masses, $A$- and
$B$-terms remain vanishing. At the messenger scale,
\begin{eqnarray}
 M_{\rm mess}\ ,
\end{eqnarray}
the messenger fields decouple. The threshold corrections (i.e., gauge
mediation effects) contribute to the gaugino masses, sfermion masses and
also the Higgs masses squared. Those are calculable with a single
parameter,
\begin{eqnarray}
 \bar M \equiv {1 \over (4 \pi)^2} {F_S \over \langle S \rangle}\ ,
\end{eqnarray}
as we can read off from Eq.~(\ref{eq:set-up}). This parameter controls
the overall scale of the supersymmetry breaking parameters.
For example, the gluino mass is $M_3 = g_3^2 \bar
M$~\cite{Dine:1993yw,Dine:1994vc,Dine:1995ag}. The $A$- and $B$-terms
are still vanishing (up to higher order loop
corrections~\cite{Rattazzi:1996fb}) at the messenger scale, but the RG
evolution below the messenger scale generates those through one-loop
diagrams.

All the soft supersymmetry breaking parameters at the electroweak scale
can be expressed in terms of these four parameters, $m_H^2$, $\mu$,
$M_{\rm mess}$, and $\bar M$, by the procedure described above. One
combination of the parameters should be fixed by the condition for the
electroweak symmetry breaking, i.e., $M_Z = 91.2$~GeV. We take the
$m_H^2$ parameter as an output of the calculation. The model parameters
are now defined by $(\mu, M_{\rm mess}, \bar M)$. Here we take the
running $\mu$ parameter at the scale $M_{\rm SUSY} \equiv ( m^2_{{\tilde
t}_L} m^2_{{\tilde t}_R} )^{1/4}$ as an input parameter.

\begin{figure}[t]
\begin{center}
  \includegraphics[width=7.5cm]{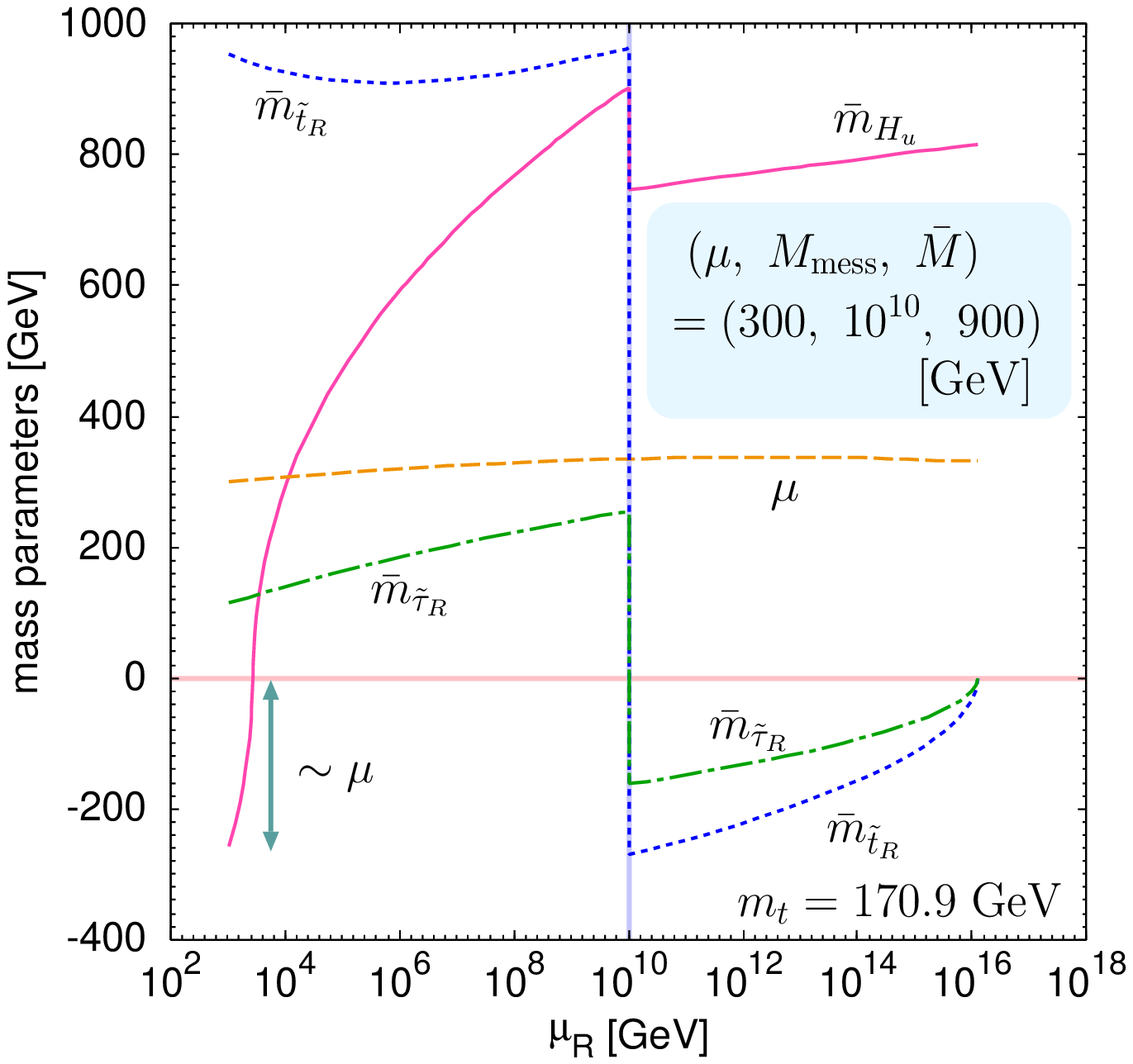}
  \includegraphics[width=7.5cm]{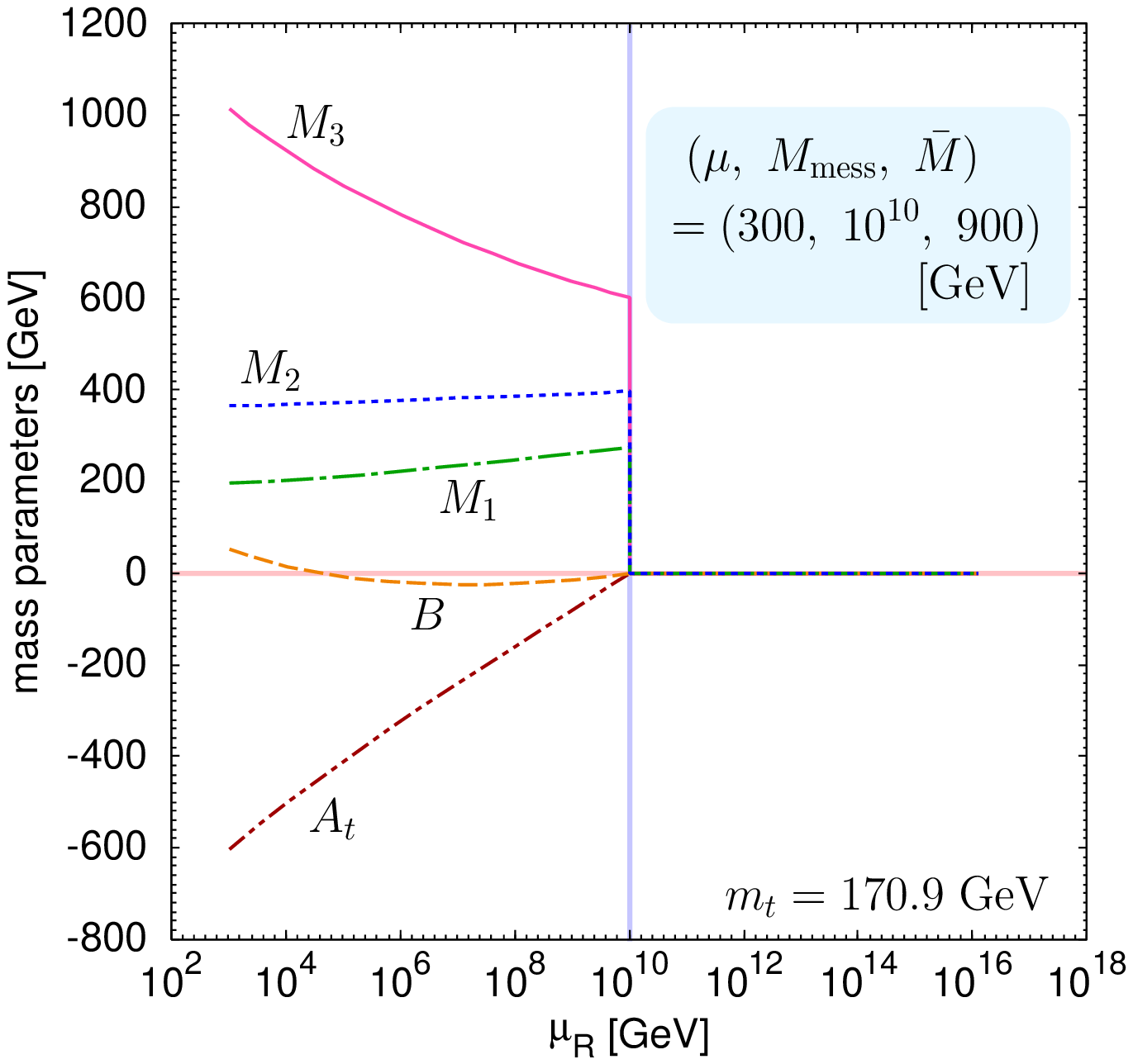}
\end{center}
\caption{The RG evolution of the supersymmetry breaking parameters. RG
 equations at one-loop level are used.  A parameter set $(\mu, M_{\rm
 mess}, \bar M) = (300, 10^{10}, 900)$~[GeV] is chosen.  The left panel
 shows the evolution of the soft masses for $\tilde t_R$ (dotted),
 $\tilde \tau_R$ (dot-dashed), and $H_u$ (solid). The $\bar m_X$
 parameter is defined by $\bar m_X \equiv {\rm sgn}(m_X^2)
 |m_X^2|^{1/2}$ for each chiral superfield $X$. The evolution of the
 $\mu$-parameter (dashed) is also shown. Negative contributions for
 $\bar m_{\tilde t}$ and $\bar m_{\tilde \tau}$ above the messenger
 scale comes from the one-loop contribution through the Yukawa
 interactions. Threshold effects (gauge mediation) at the messenger
 scale contribute to sfermion and the Higgs mass parameters. The
 $m_{H_u}^2$ parameter is driven to a negative value by the stop-loop
 diagrams. In the right panel, gaugino masses, $A$-, and $B$-parameter
 are shown. The gaugino masses are generated at the messenger scale, and
 induces $A$- and $B$-terms by the one-loop running. For the phase
 convention of $A$- and $B$-terms, we have used the one defined in
 Ref.~\cite{Skands:2003cj}.}
 \label{fig:rge}
\end{figure}

We show in Fig.~\ref{fig:rge} an example of the RG evolution of soft
supersymmetry breaking parameters for $(\mu, M_{\rm mess}, \bar M) =
(300~{\rm GeV}, 10^{10}~{\rm GeV}, 900~{\rm GeV})$. The horizontal axis $\mu_R$
is the RG scale.
We have used the top quark mass, $m_t =
170.9$~GeV~\cite{unknown:2007bx}. The constraint from the electroweak
symmetry breaking fixes the $m_{H}^2$ parameter to be $(817~{\rm
GeV})^2$.
The choice of parameters is motivated by the discussion in the last
section. The positive value of $m_{H}^2$ and a relatively small value of
$\mu (M_{\rm GUT})$ compared to $\sqrt {m_H^2}$ are realized with this
set of parameters. The lightest Higgs boson mass is calculated to be
115~GeV. We will use this set of parameters in a collider study in
Section~\ref{sec:lhc}.

In the left panel of Fig.~\ref{fig:rge}, scalar masses and the
$\mu$-parameter are plotted. We have defined mass parameters $\bar m_{X}
\equiv {\rm sgn} (m_X^2) | m_X^2 |^{1/2}$ for each scalar mass parameter
$m_X^2$. Several interesting things are happening here. With non-zero
positive values of $m_{H}^2$ the Yukawa interactions induces negative
masses squared for sfermions in the third generation.  The positive
values are motivated by the positivity of the function $g(x)$ in
Eq.~(\ref{eq:g-func}).
The negative contributions to the sfermion masses are compensated by the
positive contributions from the gauge mediation effects at the messenger
scale. This behavior of the RG evolution gives smaller values of the
stau mass, $m_{\tilde \tau_R}^2$, at the electroweak scale compared to
those in the conventional gauge mediation scenarios. The impact on the
collider physics of this effect will be discussed in
Section~\ref{sec:lhc}.

As is clear from Eq.~(\ref{eq:ewsb}) the value of the $\mu$-parameter is
approximately obtained by $\mu \simeq - \bar m_{H_u} |_{\mu_R \sim {\rm
TeV}}$. Therefore, we can easily understand from this figure that the
value of $\mu$ is smaller for larger initial values of $m_{H}^2$. On the
other hand, the stau mass is also smaller for large $m_H^2$ because it
receives more negative contributions. This correlation between the
$\mu$-parameter (the Higgsino masses) and the stau masses is an
interesting prediction of the model. We will discuss this relation more
in subsection~\ref{sec:higgsino-stau}.

The evolution of gaugino masses, $A$- and $B$-terms are shown in the
right panel of Fig.~\ref{fig:rge}. We have used the sign convention of
those parameters defined in Ref.~\cite{Skands:2003cj}. Those $R$-charged
parameters are generated at the messenger scale by gauge mediation
effects. We see a peculiar behavior of the $B$-term. It starts from zero
at the messenger scale, goes to negative once and flips its sign to
positive later. The negative contribution comes from a loop diagram with
the SU(2) gaugino, and a stop-loop diagram with the $A_t$ term gives a
positive contribution. The behavior can be understood by the fact that
the $A_t$ term starts at zero but its absolute value becomes larger at
low energy.
The positivity of the $B$-term remains to be true unless the messenger
scale is extremely low such as $M_{\rm mess} \lesssim 10^5$~GeV.
In this case, the two physical signs are predicted to be
\begin{eqnarray}
 {\rm sgn} (M_{i} \mu (B \mu)^*) = +1\ ,\ \ \ 
 {\rm sgn} (M_{i} A_t^*) = -1\ \ \ \ (i = 1,2,3)\ .
\end{eqnarray}
Unlike other models used in literatures for collider studies, there is
no choice of these signs.
The former predicts a positive sign for the supersymmetric contributions
to the anomalous magnetic moment of muon. Also, the latter determines
the sign of the chargino-loop contribution to the amplitude of the $b
\to s \gamma$ decay to be opposite to the one from loop diagrams with
the charged Higgs boson. Interestingly, both of these signs are
preferred by the experimental constraints on these quantities.

\subsection{Electroweak symmetry breaking}

\begin{figure}[t]
\begin{center}
  \includegraphics[height=7.5cm]{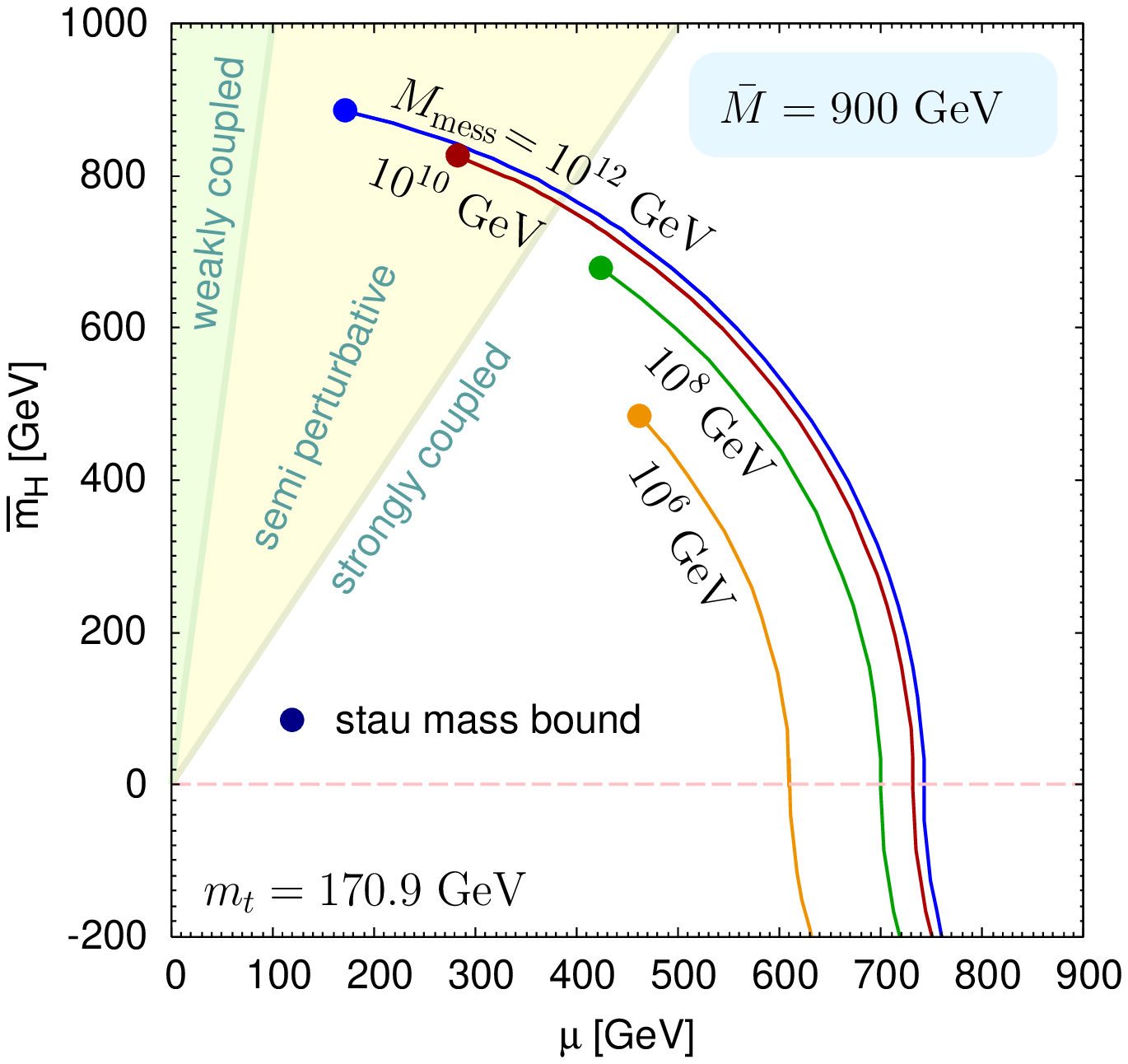}
  \includegraphics[height=7.5cm]{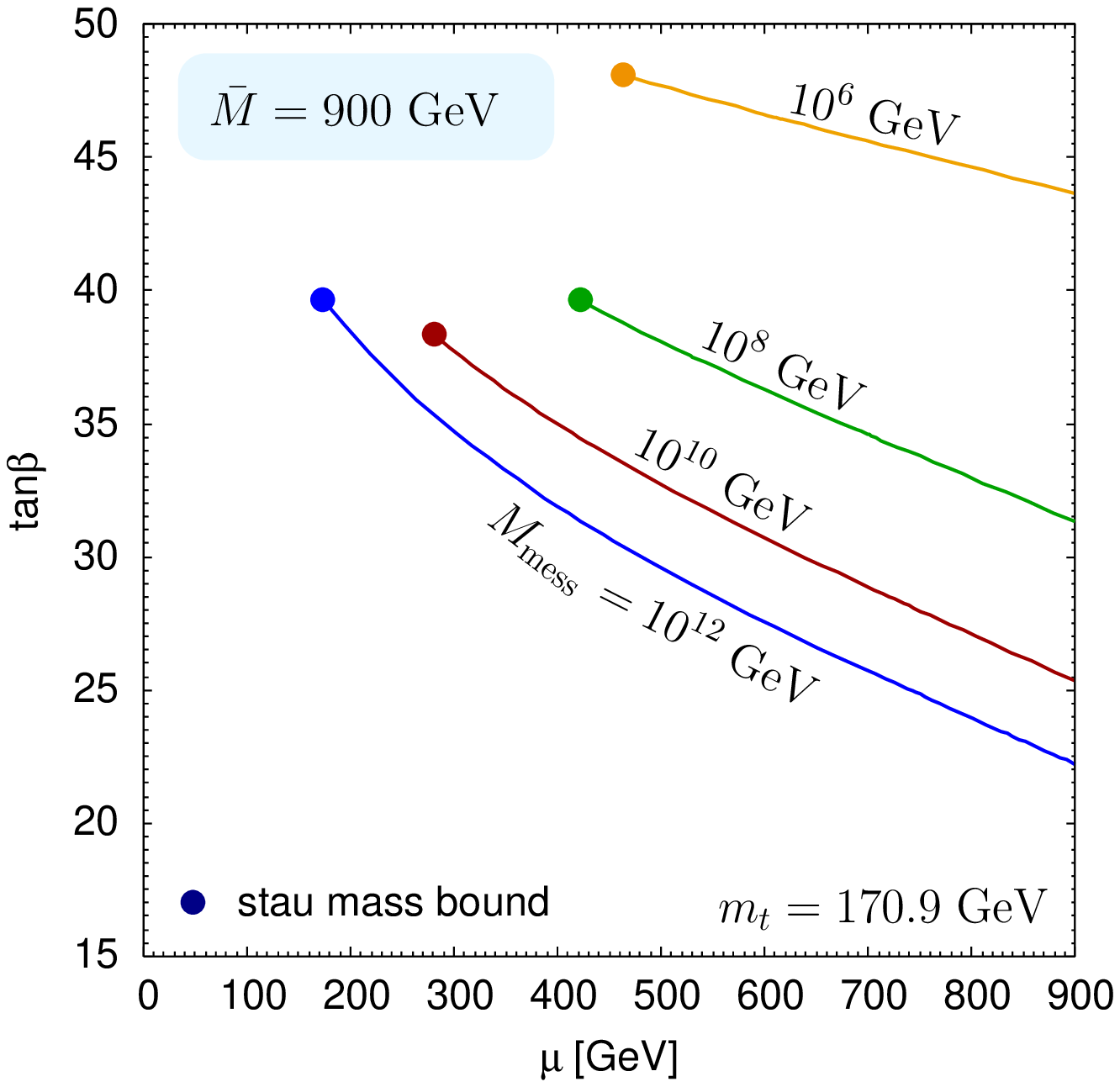}
\end{center}
\caption{The corresponding value of $\bar m_H$ (left) and $\tan \beta$
 (right) to the input parameter $\mu$ ($\mu (M_{\rm SUSY})$). We set the
 overall scale $\bar M = 900$~GeV.  For different values of $\bar M$ we
 can obtain approximate relations by rescaling the axes. Curves for
 messenger scales $M_{\rm mess} = 10^6$, $10^8$, $10^{10}$, and
 $10^{12}$~GeV are shown. The curves are terminated by the mass bound of
 stable staus $m_{\tilde \tau_1} > 98$~GeV~\cite{Abbiendi:2003yd}. Small
 values of $\mu^2/\bar m_H^2$ are predicted if the UV theory is weakly
 coupled. A rough classification of `weakly coupled', `semi
 perturbative', and `strongly coupled' is indicated.
  } \label{fig:tanb}
\end{figure}

It is highly non-trivial whether we can have successful electroweak
symmetry breaking with the limited number of parameters. We demonstrate
here that the correct $Z$-boson mass can be obtained without spoiling
the perturbativity of the Yukawa interactions for the top and bottom
quarks up to the GUT scale.

The left panel in Fig.~\ref{fig:tanb} shows the value of $\bar m_H$
(defined at the GUT scale) required by the correct electroweak symmetry
breaking with respect to the running $\mu$-parameter at $M_{\rm
SUSY}$. Curves for different messenger scales $M_{\rm mess}$ are shown.
The $\bar M$ parameter is fixed to be 900~GeV. For other values of $\bar
M$, say $x \bar M$ with an arbitrary positive number $x$, we can obtain
curves, as a good approximation, by rescaling the both axes by the
factor of $x$.
%
%
As is clear from the behavior of the RG evolution shown in
Fig.~\ref{fig:rge}, larger values of $\bar m_H$ are necessary for having
smaller values of $\mu$.
Each lines are terminated at some small value of $\mu$, where the stau
becomes too light ($m_{\tilde \tau_1} < 98$~GeV~\cite{Abbiendi:2003yd})
due to the negative contribution from the running between the GUT scale
and the messenger scale. This negative contribution is significant only
when $\tan \beta$ is large. It is indeed the case as we will see later.

As we discussed in the previous section, the UV completion of the theory
above the GUT scale suggests that the Higgs fields do not get strongly
coupled.
The discussion is based on the requirement of the $O(1)$ Yukawa coupling
constant for the top quark.
This indicates a (small) hierarchy between the $m_H$-parameter and the
$\mu$-parameter at the GUT scale since the ratio $\mu^2 / m_H^2$ turns
out to be the loop-expansion parameter. We indicate in the plot the
region with $0 \leq \mu^2/\bar m_H^2 < 1/100$ to be `weakly coupled',
$1/100 \leq \mu^2/ m_H^2 < 1/4$ to be `semi perturbative', and $\mu^2 /
m_H^2 \geq 1/4$ or $m_H^2 < 0$ to be `strongly coupled' for
illustration. (There is no concrete meaning in precise locations of
border lines.)\footnote{Precisely speaking, the hierarchy is predicted
for the ratio of $\bar m_H$ and the $\mu$ parameter at the GUT scale
whereas what is plotted is the $\mu$-parameter at $M_{\rm
SUSY}$. However, the RG running of the $\mu$ parameter changes its value
only by $O(10\%)$ (see Fig.~{\ref{fig:rge}}).}
We only find solutions with small $\mu$ compared to the overall scale
$\bar M$ if we impose (semi) perturbativity. The light Higgsino is,
therefore, one of the predictions of the model.
Also, a high messenger scale, such as $M_{\rm mess} \gtrsim 10^8$~GeV is
required to be in that region.

There is a possibility that the Higgs fields are fully strongly coupled
and the top Yukawa coupling is generated by some strong dynamics by
making the top quark also involved in the strong sector. Since there is
no stringent constraint for flavor violation in the top-quark physics,
this is not a dangerous assumption although the embedding to GUT models
may be more difficult. In this case, the stop mass squared at the GUT
scale becomes an additional parameter of the model. We do not consider
this modification of the framework in this paper.

We show in the right panel the predicted values of $\tan \beta$. Again,
we take $\bar M = 900$~GeV. The rescaling of the horizontal axis gives a
curve for different values of $\bar M$ with a good accuracy. For small
values of $\mu$ preferred by UV completions, $\tan \beta \sim 30-40$ is
predicted.
The values in this range do not cause a Landau pole of the Yukawa
coupling constants below the GUT scale. Therefore, the perfectly
consistent electroweak symmetry breaking is achieved without any
extension of the model.
Large values of $\tan \beta$ are due to the fact that the $B \mu$ term
is generated only through radiative corrections below the messenger
scale~\cite{Rattazzi:1996fb}. The Yukawa coupling constant for the tau
lepton is large for large $\tan \beta$, which affects the running of
$m_{\tilde \tau_R}^2$.

\subsection{Light {\boldmath $\tilde \tau$} and light Higgsino}
\label{sec:higgsino-stau}

\begin{figure}[t]
\begin{center}
  \includegraphics[height=7.5cm]{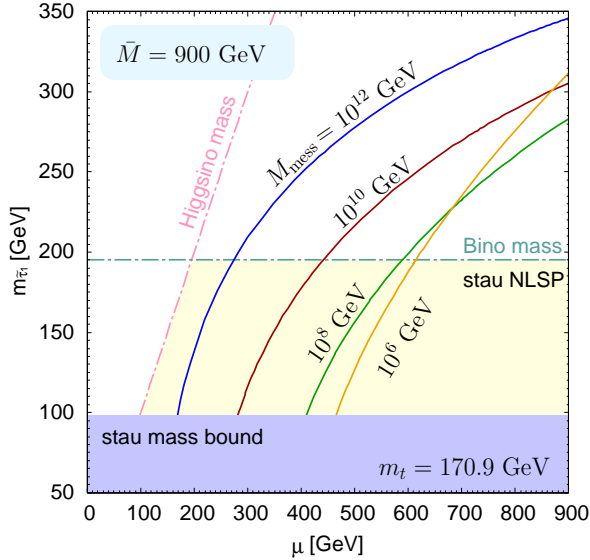}
\end{center}
\caption{Correlation of the $\mu$ parameter and the lighter stau
 mass. The overall scale $\bar M = 900$~GeV is set. Below the lines of
 the Bino mass and the Higgsino mass, the stau is the NLSP.  We can
 clearly see the positive correlation.}  \label{fig:stau}
\end{figure}

The characteristic RG evolution of the supersymmetry breaking parameters
in this framework provides interesting predictions in low energy
physics.
In particular, the correlation of the $\mu$-parameter and the stau mass
gives a large impact on collider physics.

The plot in Fig.~\ref{fig:stau} shows the correlation for different
values of $M_{\rm mess}$. For small values of $\mu$, the stau is lighter
than the Bino and Higgsinos. Therefore, for values of $\mu$ in the
(semi) perturbative region (see left panel of Fig.~\ref{fig:tanb}) there
is a big chance that the stau becomes the NLSP (remember that the
gravitino is the LSP). The collider physics of this region will be very
different from the scenario with the neutralino NLSP as there is no
missing $E_T$ associated with escaping neutralinos. (The lifetime of the
stau is typically of $O(1000)$~seconds with which the stau is regarded
as a completely stable particle for the time scale of collider
experiments.) Also, the light Higgsino changes the pattern of cascade
decays of supersymmetric particles. We will discuss in the next section
the overall feature of this scenario at the LHC and demonstrate a way of
confirming/excluding the framework.

\section{LHC signatures}
\label{sec:lhc}

\setcounter{footnote}{0}

The theoretical success of the sweet spot supersymmetry motivates us to
consider what will be the experimental signatures at the LHC
experiments. We show in this section that there are several unique
features. We present a way of confirming/excluding the model in the case
where the lighter stau is the NLSP.

\subsection{Overview of supersymmetric events with {\boldmath
  $\tilde \tau$} NLSP}

As we have seen in the last section, it is plausible that the lighter
stau is the NLSP. A small value of the $\mu$-parameter is a natural
consequence of UV physics, and that makes $\tilde \tau$ light through
the RG evolution. If it is the NLSP, the lifetime of stau is of
$O(1000)$~seconds with our assumption of the $O(1)$~GeV gravitinos.
The LHC signals with such a long-lived stau will be quite different from
ones with the usual assumption of the neutralino LSP.
There have been many studies on collider signatures for the quasi-stable
$\tilde \tau$-NLSP scenario, for example, in
\cite{Drees:1990yw}-\cite{Cakir:2007xa}.  We demonstrate here
reconstruction of model parameters with $\tilde \tau$ NLSP at the LHC
experiments.\footnote{Recent studies on the lifetime measurement of the
long-lived charged NLSP in the collider experiments show that it is
possible to determine the gravitino mass in some range of the parameter
region~\cite{Hamaguchi:2004df}-\cite{Hamaguchi:2006vu} although those
proposals require an extra experimental set-up to collect the charged
NLSP.}

\begin{table}
\begin{center}
\begin{tabular}{|l|l||l|l|}
\hline\hline
 ${\tilde{g}}$        & 1013       &  ${\tilde{\nu}_L}$  & 543   \\
 ${\chi_{1}^{\pm}}$     & 270      & ${\tilde{t}_1}$  & 955   \\
 ${\chi_{2}^{\pm}}$        & 404     & ${\tilde{t}_2}$  & 1177\\
 ${\chi_{1}^{0}}$        & 187      & ${\tilde{b}_1}$  & 1128\\
 ${\chi_{2}^{0}}$        & 276      & ${\tilde{b}_{2}}$  & 1170   \\
 ${\chi_{3}^{0}}$        & 307      & ${\tilde{\tau}_{1}}$  & 116   \\
 ${\chi_{4}^{0}}$        & 404     & ${\tilde{\tau}_2}$  & 510   \\
 ${\tilde{u}_{L}}$       & 1352    & ${\tilde{\nu}_{\tau}}$  & 502   \\
 ${\tilde{u}_{R}}$     & 1263      & ${h^{0}}$  & 115   \\
 ${\tilde{d}_{L}}$     & 1354      & ${H^{0}}$  & 770   \\
 ${\tilde{d}_{R}}$     & 1251      & ${A^{0}}$  &    765 \\
  ${\tilde{e}_{L}}$     & 549      & ${H^{\pm}}$  & 775    \\
 ${\tilde{e}_{R}}$     & 317      & $\tilde G$  &   0.5 \\
\hline\hline
\end{tabular}
\end{center}
\caption{Masses of superparticles and Higgs bosons in GeV for our
benchmark point, $\mu = 300$\,GeV, $M_{\rm mess} = 10^{10}$\,GeV and
$\bar{M} = 900$\,GeV.  The gravitino mass is fixed to account for the
observed dark matter density (see Eq.\,(\ref{eq:omega})).  Here, the
masses of the squarks and sleptons of the second generation are omitted,
since they are equal to the ones of the first generation.  We use the
notation for the superparticles and Higgs bosons in the MSSM in
Ref.~\cite{Eidelman:2004wy}.  } \label{tab:spectrum}
\end{table}


We select the following benchmark point for the collider study:
\begin{eqnarray}
\mu = 300~{\rm GeV}\ , \ \ M_{\rm mess} = 10^{10}~{\rm GeV}\ ,\ \ \bar M
 = 900~{\rm GeV}\ .
\end{eqnarray}
This set represents the most theoretically motivated region of the
parameter space as we have discussed before.
As we can see in Fig.~\ref{fig:stau} the NLSP is the stau with this set
of parameters.
We have calculated the spectrum by solving RG equations at one-loop
level. The running parameters at the scale $M_{\rm SUSY} \equiv
(m^2_{\tilde t_L} m^2_{\tilde t_R})^{1/4} = 1053$~GeV have been used for
the calculation of the spectrum. We have ignored the QCD finite
corrections at the low energy threshold, which would amount to about
10\%.
In Table~\ref{tab:spectrum}, we listed masses of superparticles and
Higgs bosons. We used $m_t = 170.9$~GeV. The stau mass is 116~GeV, and
its lifetime is calculated to be 3000~seconds with the gravitino mass
determined by the dark matter density, $m_{3/2} = 500$~MeV.
The running gaugino mass parameters at $M_{\rm SUSY}$, $M_i = g_i^2 \bar
M$, are $M_1 = 195$~GeV, $M_2 = 364$~GeV, and $M_3 = 1013$~GeV. The
lightest neutralino $\chi_1^0$ is, therefore, mostly the Bino,
$\chi_2^0$ and $\chi_3^0$ mainly consist of the Higgsino components, and
the Wino is the heaviest, $\chi_4^0$. The lighter and heavier charginos,
$\chi_1^\pm$ and $\chi_2^\pm$, are mainly the Higgsino and the Wino,
respectively.
The gluino and squark masses are about $1\,$TeV.  
The Higgs boson mass, 115~GeV, is calculated by using a one-loop
effective potential with taking into account leading two-loop
corrections by appropriately choosing a renormalization scale for the
running top quark mass which appears in the effective
potential~\cite{Drees:1991mx}.
Similar values, $114-115$~GeV, are obtained by using publicly available
codes~\cite{Heinemeyer:1998yj,Djouadi:2002ze}.

The total cross section of the superparticle production at the benchmark
point is $1.4$\,pb for the center-of-mass energy of the LHC. The cross
section is dominated by pair productions of $\tilde{g}\tilde{g}$,
$\tilde{q}\tilde{g}$, and $\tilde{q}\tilde{q}$.
The subsequent decays of these colored particles generate hard jets and
other supersymmetric particles such as neutralinos and charginos. The
decays of these non-colored superparticles, in the end, produce two
quasi-stable $\tilde \tau_1$'s for each supersymmetric event. Most of
the stau pairs escape a detector and leave two charged tracks.

\begin{figure}[t]
\begin{center}
  \includegraphics[height=2.8cm]{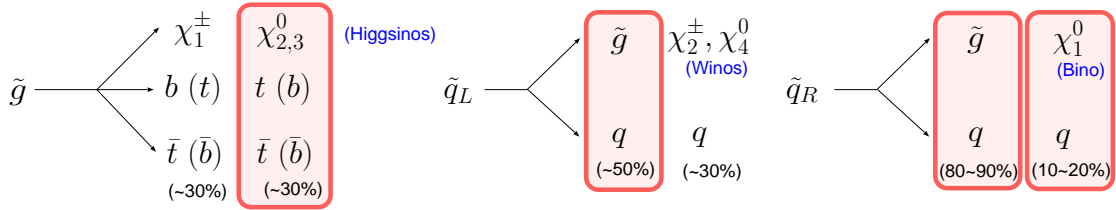}
\end{center}
\caption{The dominant decay modes of $\tilde{g}$ and $\tilde{q}$.  The
percentages show the branching ratios of each modes.  The shaded modes
are relevant for the analysis of the reconstruction of the neutralino
masses.}
 \label{fig:decay-primary}
\end{figure}

\begin{figure}[t]
\begin{center}
  \includegraphics[height=2.49cm]{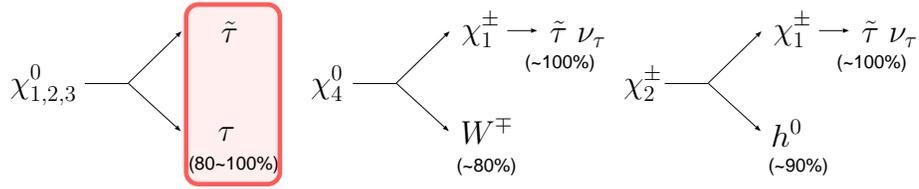}
\end{center}
\caption{The dominant decay modes of $\chi^{0}$ and $\chi^{\pm}$.  The
shaded modes are relevant for the analysis of the reconstruction of the
neutralino masses.}
 \label{fig:decay-neutralino}
\end{figure}

The decay cascades start with the the decays of $\tilde{g}$ and
$\tilde{q}$ as shown in Fig.~\ref{fig:decay-primary}.  We have used
ISAJET~7.69~\cite{Paige:2003mg} to calculate the branching ratios.
Since the gluino is lighter than squarks, it decays into a neutralino or
a chargino through three-body decay modes.
The dominant channel is the decay into a pair of third generation quarks
and a Higgsino, $\chi_1^\pm$ or $\chi_{2,3}^0$, through the Yukawa
interaction of the top quark.
The main decay mode of the squarks are $\tilde q \to \tilde g + q$,
followed by the gluino decay.
Therefore, for each supersymmetric event, many hard jets are
produced. Especially, a significant number of $b$-jets are produced by
the gluino decays (and also by the subsequent decays of the top
quarks). This is an interesting feature of the model, but at the same
time, the large number of jets makes it difficult to trace back the
decay chains in the actual analysis because it is hard to specify which
jet is the one originated from a particular decay process.

At the end of the decay chain, three lighter neutralinos (Bino and
Higgsinos) decay into $\tilde\tau_{1}$ and $\tau$
(Fig.~\ref{fig:decay-neutralino}).
The heaviest neutralino and the heavier chargino (Winos) decay into 
$\chi_1^\pm$ which in turn decays into a stau and a neutrino.
We can measure the mass and the momentum of $\tilde{\tau}_{1}$'s in the
final state by using information on the charged tracks in the muon
system.
With the full reconstruction of the four momentum of the staus, three of
neutralino masses can in principle be measured as the invariant mass of
$\tilde{\tau}_{1}$ and $\tau$.

Sleptons except for $\tilde \tau_1$ do not appear in the decay
cascades. That means we do not have clear lepton signals such as two
opposite-sign same-flavor leptons from the $\chi_2^0$ decay often used
as a tool for precision measurements~\cite{Hinchliffe:1996iu}. What we
typically have are a lot of third generation quarks ($b$-jets) and
leptons ($\tau$-jets) in the final states. It is again interesting but
not exciting situation for measurements of the mass spectrum.

\subsection{Reconstruction of neutralino masses}

We demonstrate here a way of reconstructing the neutralino masses by
using the decay modes $\chi_{1,2,3}^0 \to \tilde \tau_1 \tau$, and show
that it is possible to determine the three model parameters.

There have been studies on the mass reconstruction of the stau and the
neutralinos in stau NLSP scenarios.
Especially, in Ref.~\cite{Ambrosanio:2000ik}, a detailed study of the
stau mass measurement is performed including detector effects. The
method is measuring the velocity ($\beta$) of the stau by the
time-of-flight and the momentum ($p_{\tilde \tau_1}$) from the track. We
can then calculate the mass by the formula,
\begin{eqnarray}
 m_{\tilde \tau_1} 
= {p_{\tilde \tau_1}  \over \beta \gamma}\ .
\label{eq:stau-mass}
\end{eqnarray}
It is concluded that the stau mass can be measured with the accuracy of
at least about $100$~MeV for $m_{\tilde \tau_1} \simeq 100$~GeV.
The standard model background is estimated in the paper. Since the staus
with high velocities are indistinguishable from muons, a tight selection
cut on the velocity, $\beta \gamma < 2.2$, is imposed in the
analysis. 
We reproduced the analysis of the stau mass measurement and found that it
is indeed possible to measure it with a very good precision at the
benchmark point.
We follow the selection cuts proposed in this paper in the analysis of
the reconstruction of neutralino masses as well.

The reconstruction of the neutralino masses has been discussed in
Ref.~\cite{Hinchliffe:1998ys} and also in a recent
paper~\cite{Ellis:2006vu}.
Once we determine the stau mass, we know the four-momentum of the stau
on event-by-event basis. By combining with the four-momentum of $\tau$,
we can extract the neutralino masses.
There are two things needs to be done in the analysis.
Since every supersymmetric event contains two staus in the final state,
there are two candidate staus to be combined with for each $\tau$.
The invariant mass coincides with a neutralino mass only if we choose a
correct combination. We need a strategy to select the correct one.
The other thing is that we do not know the four-momentum of $\tau$. The
$\tau$ particle decays inside the detector and an invisible neutrino in
the decay products always carries away some amount of energy.

In the analysis in Ref.~\cite{Hinchliffe:1998ys}, it has been assumed
that the selectron and the smuon are also quasi-stable. Therefore, by
selecting events with only one stau there is no combinatorial
background. A method of fully reconstructing the four-momentum of $\tau$
has been discussed in the paper. If the direction of missing momentum is
aligned with that of the $\tau$-jet, it can be identified with the
neutrino. By adding missing momentum to the $\tau$-jet momentum, four
momentum can be reconstructed. Even without trying to reconstruct the
four-momentum of $\tau$, the authors found that there appear sharp edges
at the neutralino masses in the distribution of the $\tilde \tau - \tau$
invariant mass by using hadronic decay modes of $\tau$.

We follow this endpoint analysis with hadronically decayed $\tau$'s for
the measurement of the neutralino masses.
Since there are always two staus in the final state in contrast to the
data set analyzed in Ref.~\cite{Hinchliffe:1998ys}, we need to consider
a way of reducing the combinatorial background. Also, with updated
experimental bound on the Higgs boson mass, the overall scale of the
supersymmetric particles should be higher than the points studied in
Ref.~\cite{Hinchliffe:1998ys}. 
This significantly reduces the
statistics, which are essential for the endpoint analysis.
Also, with the cut on the stau velocity discussed above, the number of
signal events are again significantly reduced with the relatively light
stau, 116~GeV, at the benchmark point. It is, therefore, non-trivial
whether the method works in our model.

It has been recently proposed to reconstruct the energy of neutrinos by
decomposing the missing momentum into two directions of $\tau$'s which
we know from directions of leptons or $\tau$-jets~\cite{Ellis:2006vu}.
Also, by using information of the electric charge of staus and leptons
in the final state, we can select the correct combination for the events
with $\tilde \tau^+$ and $\tilde \tau^-$.
However, in our model, there are often other neutrinos in the events,
e.g., from chargino decays and top-quark decays. The decomposition fails
in the presence of such neutrinos. 
In the leptonic decays of $\tau$, neutrinos tend to carry away majority
of $\tau$ energy due to the kinematics and the $V-A$ current structure
of the weak interaction.
Therefore, all of the uncertainties in the reconstruction of the
neutrino momentum reflect to the mass measurement.
For the background rejection, the authors have used a loose cut on the
stau velocity, $\beta \gamma < 6$, and they checked that it is enough to
reject the standard model background from mis-identifications of muons
as staus.
However, with a light stau at our benchmark point, it is non-trivial
whether such a loose cut can effectively suppress the background.
The overlap between the signal and background regions in the
$(\beta\gamma,m_{\tilde \tau})$ plane gets larger for a light stau.
Once we impose a tighter cut, $\beta \gamma < 2.2$, we could not obtain
enough number of events for the analysis.
Therefore, we do not pursue this direction.

In the following, we discuss a method of the simulation and selection
cuts for reducing the standard model background. The actual analysis
will be presented in \ref{sec:analysis}.

\subsubsection{Event generation and selection cuts}
We have generated 42,900 events of the superparticle productions in
proton-proton collision at the LHC energy by using a event generator
HERWIG~6.50\,\cite{Corcella:2002jc} with the MRST\,\cite{Martin:1998np}
parton distribution function.  The number of the events is equivalent to
the integrated luminosity of $30\,{\rm fb}^{-1}$.
We have used the package TAUOLA~2.7 for $\tau$
decays~\cite{Jadach:1993hs}.

For a detector simulation, we have used a package
AcerDET-1.0\,\cite{Richter-Was:2002ch}, which is a fast simulator for
high $p_T$ physics at the LHC.
The AcerDET program identifies isolated leptons, isolated photons and
isolated jets out of the final state of each generated event.
The cluster selections and the smearing of the
four-momentum of leptons, photons and jets are implemented.  Leptons and
photons are considered to be isolated if they are far from other
clusters by $\Delta R > 0.4$, and the transverse energies deposited in
cells in a cone $\Delta R= 0.2$ around the cluster are less than
$10$\,GeV.  A cluster is recognized as a jet by a cone-based algorithm
if it has $p_{T} > 15$\,GeV in a cone $\Delta R = 0.4$.  The package
also implements a calibration of jet four-momenta using a flavor
independent parametrization, optimized to give a proper scale for the
di-jet decay of a light Higgs boson. 
Each jet is labeled either as a light jet, $b$-jet, $c$-jet or
$\tau$-jet, using information of the event generators.  We have used
default values of the parameters for clustering, selection, isolation,
calibration, and labeling processes.  For the $\tau$-jet identification,
we further implement $\tau$-tagging efficiency of 50\% per a
$\tau$-labeled jet.

A parametrization of the resolution of the stau velocity is shown in
Ref.~\cite{Ambrosanio:2000ik}:
\begin{eqnarray}
 {\sigma (\beta) \over \beta } = 2.8\% \times \beta.
\end{eqnarray}
We smeared the velocity by using this resolution. The same resolution is
assumed for the measurement of the muon velocity.

For the smearing of the stau momentum, we have used the momentum
resolutions shown in Ref.~\cite{Ellis:2006vu}\footnote{According to the
paper, the original study has been done by G.~Polesello and A.~Rimoldi,
in ATLAS Internal Note ATL-MUON-99-06, but it is not publicly
available.}; one from the sagitta measurement error,
\begin{eqnarray}
\frac{\sigma(p_{\tilde \tau_1} )}{p_{\tilde \tau_1} } 
= 0.0118\% \times (p_{\tilde \tau_1}  / {\rm GeV}),
\end{eqnarray}
one from a multiple scattering term,
\begin{eqnarray}
\frac{\sigma(p_{\tilde \tau_1} )}{p_{\tilde \tau_1} } 
= 2 \% \times \sqrt{1 + \frac{m_{\tilde\tau_{1}}^{2} }{p_{\tilde \tau_1} ^{2}} },
\end{eqnarray}
and one from the fluctuation of energy loss in the calorimeter, 
\begin{eqnarray}
\frac{\sigma(p_{\tilde \tau_1} )}{p_{\tilde \tau_1} } 
= 89\% \times (p_{\tilde \tau_1} /{\rm GeV})^{-2}
\end{eqnarray}
We have smeared the stau momentum according to these resolution width
$\sigma(p_{\tilde \tau_1} )$.  

If the measured velocity of the stau is high enough, such as $\beta
\gamma > 0.9$~\cite{Ellis:2006vu}, the stau will be identified with a
muon and can be used as a trigger. However, for slow staus, we need to
rely on other triggers.
For the simulation of the triggering, we have chosen only events passing
one of the following conditions~\cite{tdr:1999fq}: one isolated electron with
$p_{T} > 20$\,GeV, one isolated photon with $p_{T} > 40$\,GeV, two
isolated electrons/photons with $p_{T} >15$\,GeV, one muon with $p_{T}>
20$\,GeV, two muons with $p_{T}> 6$\,GeV, one isolated electron with
$p_{T} > 15$\,GeV and one isolated muon with $p_{T} >6$\,GeV, one jet
with $p_{T}>180$\,GeV, three jets with $p_{T}>75$\,GeV, and four jets
with $p_{T}>55$\,GeV.  Here, isolated electrons/photons, isolated muons
and jets must be in the central regions of pseudorapidity $|\eta| < 2.5,
2.4$, and 3.2, respectively.
Following~\cite{Ellis:2006vu}, we treated staus with $\beta \gamma >
0.9$ as muons in the simulation of triggering.

For the event selection, we require two stau candidates for each
event. Since the stau mass can be precisely determined, a stau
identification can be performed by testing if its measured mass by
Eq.~(\ref{eq:stau-mass}) is consistent with the actual mass.  For the
consistency test, we took a window of the measured velocity, $\beta_{\rm
meas}$:
\begin{eqnarray}
\beta^\prime - 0.05 < \beta_{\rm meas} < \beta^\prime + 0.05\ ,
\end{eqnarray}
where $\beta^\prime$ is a velocity calculated from the measured
momentum, $p_{\rm meas}$, by assuming the stau mass, i.e., $\beta^\prime
= \sqrt{ p_{\rm meas}^2 / ( p_{\rm meas}^2 + m_{\tilde \tau_1}^2)}$~(see
\cite{Ambrosanio:2000ik}).
To reduce the standard model background from mis-identifications of
muons as staus, we required one of the stau candidate selected above to
have $\beta \gamma < 2.2$. The transverse momentum cut, $p_T > 20$~GeV,
is also imposed.
The lower limit on the velocity $\beta \gamma > 0.4$ is imposed to
ensure the stau to reach the muon chamber.
As for the isolation of stau, we have used the same criterion with that
of the muon.

The standard model background can be further suppressed by requiring the
presence of the two charged stable particles as well as a large
effective mass~\cite{Hinchliffe:1998ys}.  The effective mass $M_{\rm
eff}$ is defined by the scalar sum of the transverse momentum of the
four leading jets $p_{T,i}$\,($i=1-4$), and the missing transverse
momentum $E_{T}^{\rm miss}$,
\begin{eqnarray}
M_{\rm eff} = p_{T,1}+p_{T,2}+p_{T,3}+p_{T,4}+E_{T}^{\rm miss}.
\end{eqnarray}
The effective mass distribution of supersymmetric events has a peak
around 1~TeV.
We impose a cut $M_{\rm eff}>800$\,GeV.
With all of the selection cuts discussed above, the standard model
background is reduced to a negligible level~\cite{Ambrosanio:2000ik}.

Events with only one $\tau$-tagged jet is selected for the
reconstruction of the neutralino mass.  We require $p_T > 40$~GeV for
the $\tau$-jet.
If we allow two $\tau$-tagged jets, the number of signal events
increases by about 10\%, but it also increases the probabity of
selecting a fake $\tau$-jet.
The effects of the fake $\tau$-jets are taken into account by assuming
the probability of the mis-tagging of non-$\tau$-labelled jets to be 1\%
per a jet~\cite{tautag}. We also discuss the significance of the
fake events by comparing with a case with a 5\% mis-tagging probability.
For simplicity, we took the tau-tagging efficiency and the mis-tagging
probability to be independent of jet $p_T$, $\eta$, and the decay modes
of $\tau$.

\subsubsection{Invariant mass analysis}
\label{sec:analysis}

With the selection cuts described above, the total number of events are
reduced to 2,000 in which we have 1,563 events with a true $\tau$-jet
and a true stau pair. With these limited statistics, we need to develop
an effective method to reduce the combinatorial background.

\begin{figure}[t]
\begin{center}
  \includegraphics[height=7.5cm]{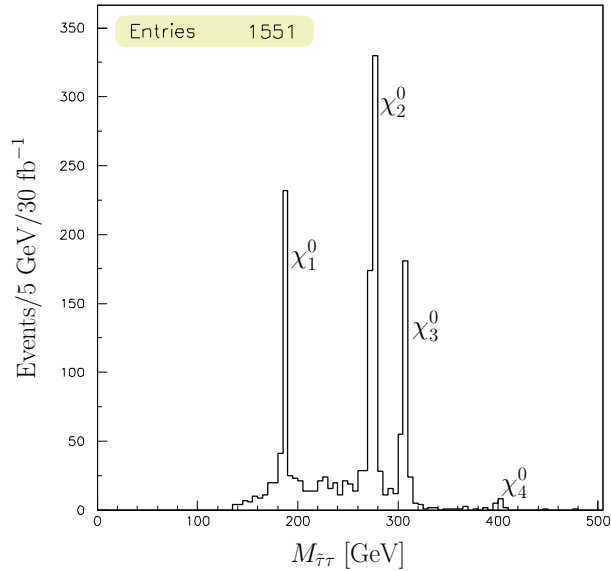}
\end{center}
\caption{The $\tilde \tau - \tau$ invariant mass distribution.
The combination with the lowest invariant mass is chosen.  The
four-momentum of $\tau$-lepton extracted from the event generator is
used.
}
 \label{fig:parton-tau}
\end{figure}

The combinatorial background can be reduced by choosing a stau which
gives the smaller value of the invariant mass $M_{\tilde \tau \tau}$ for
every $\tau$ candidate. Since the neutralinos produced by the decay of
gluinos or squarks are likely to be highly boosted, the stau and $\tau$
tend to be emitted in a similar direction, and therefore the invariant
mass is likely to be much larger than the neutralino masses if we choose
a wrong combination.
To see how it works, we show the invariant mass distribution
$M_{\tilde{\tau}\tau}$ of the combination selected by the above strategy.
We use the four-momentum of $\tau$ extracted from the event generator
(Fig.~\ref{fig:parton-tau}).  Here, no information on the charge of
$\tau$ is used.
As we see from the figure, the lowest mass combination shows peaks at
the masses of $\chi_{1,2,3}^{0}$.
The correct combinations are chosen with the probability of about 70\%.

\begin{figure}[t]
\begin{center}
  \includegraphics[height=7.5cm]{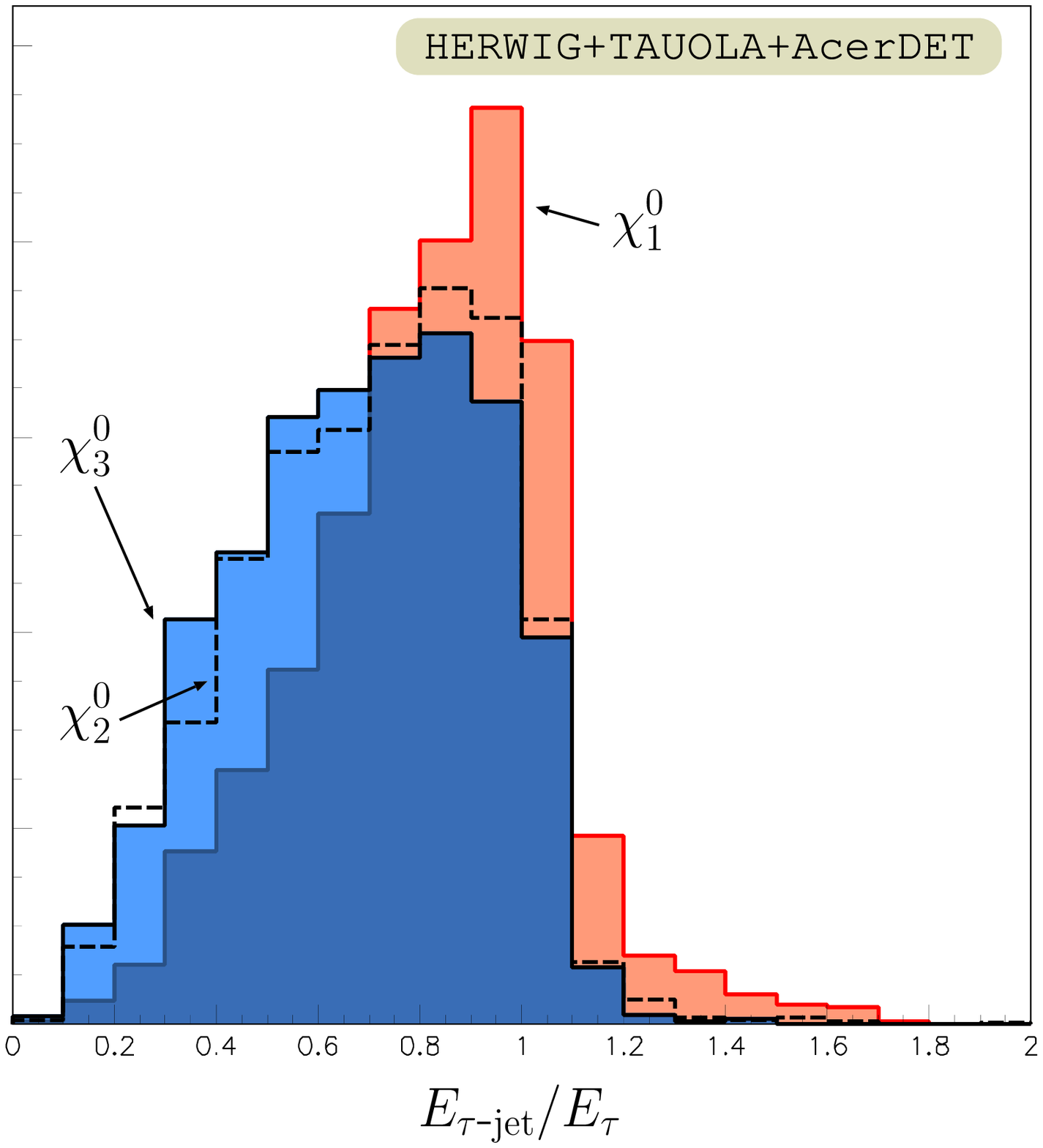}
  \includegraphics[height=7.58cm]{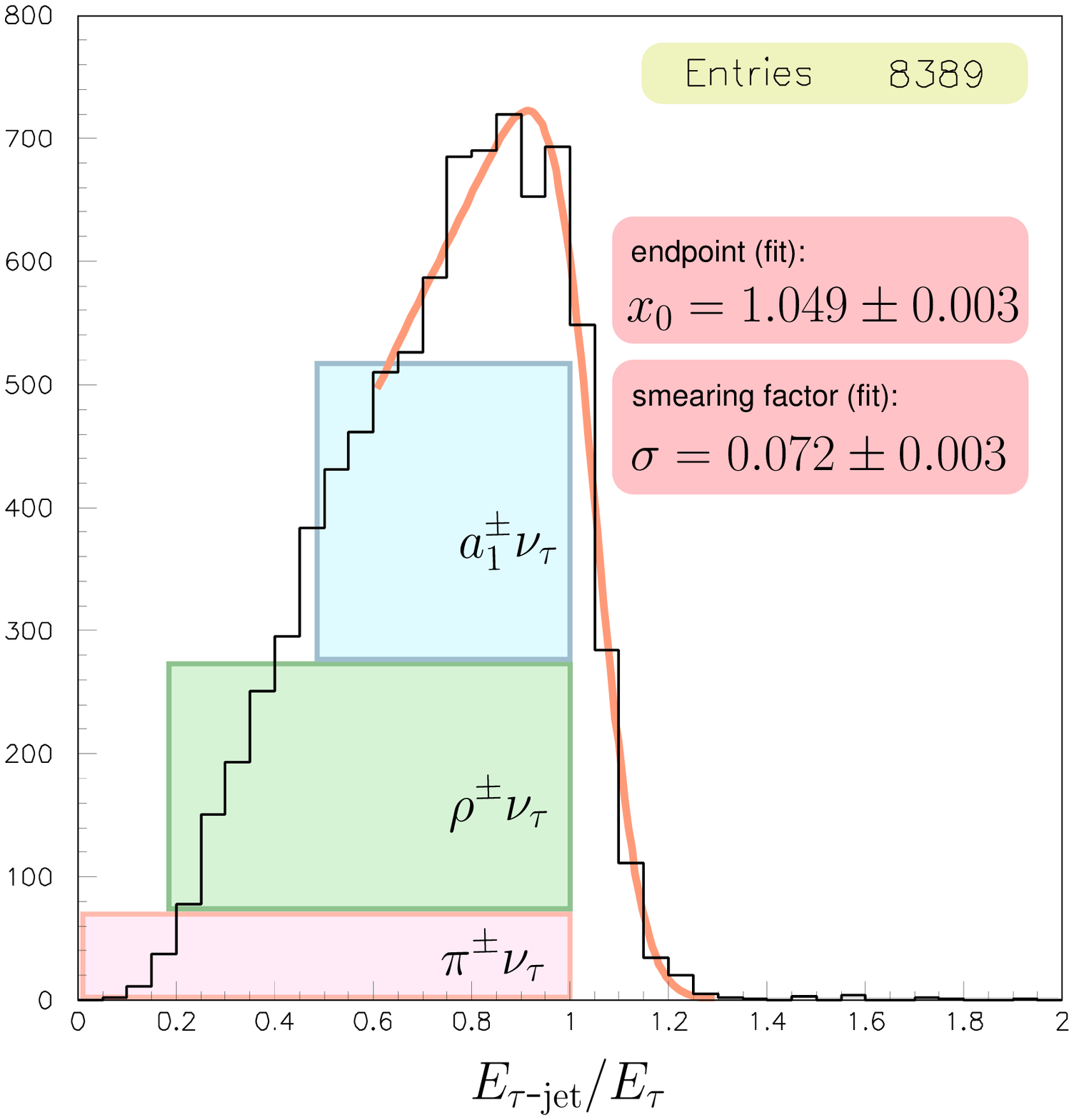}
\end{center}
\caption{The distribution of the $\tau$-jet energy fraction
$E_{\tau\mbox{-}{\rm jet}}/E_{\tau}$ in the hadronic decay modes of
$\tau$ in supersymmetric cascade decays.  In the left panel, we show the
energy fractions for $\tau$'s which originate from three species of
neutralinos, $\chi_1^0$, $\chi_2^0$ and $\chi_3^0$, respectively.  They
are rescaled so that the number of events are the same for three
neutralinos.
In the right panel, we did not distinguish the origin of $\tau$.  Shaded
histograms are the distribution of the energy fraction for the two-body
decays, $\tau\to\pi\nu$, $\tau\to\rho\nu$ and $\tau\to a_{1}\nu$
assuming stable mesons.
Energy calibration of the $\tau$-jets is performed by AcerDET.
}
 \label{fig:ejetau}
\end{figure}

In the actual experiment, however, the four-momentum of $\tau$-lepton is
not available.  Instead, the four-momentum of the $\tau$-jet from the
hadronic $\tau$ decays is available.
If we use the four-momentum of the $\tau$-jet in the $M_{\tilde \tau
\tau}$ analysis, the peaks shown in Fig.~\ref{fig:parton-tau} are
smeared by the effect of the missing energy in the $\tau$ decays.
The left panel of Fig.~\ref{fig:ejetau} shows the distribution of the
energy fraction of $\tau$-jet, $E_{\tau\mbox{-}{\rm jet}}/E_{\tau}$, in
the neutralino decays. We plotted histograms for each neutralino,
$\chi_1^0$, $\chi_2^0$ and $\chi_3^0$. We rescaled the histograms so
that the number of events are the same for each neutralino.
Energies are measured in the laboratory frame.
With the HERWIG event generator and the TAUOLA package, effects of the
polarization of $\tau$ are taken into account.
As we can see, there are sharp edges in the distribution at
$E_{\tau\mbox{-}{\rm jet}}/E_{\tau}=1$. Especially, the edge is sharper
for $\chi_1^0$ compared to $\chi_2^0$ and $\chi_3^0$. This can be
understood as an effect of the polarization of $\tau$.\footnote{We thank
L.~Dixon for pointing out the possibility of having polarization
effects.} Since the stau is mostly right-handed, the chirality of $\tau$
from the neutralino decay is right-handed (left-handed) if the
neutralino is gaugino-like (Higgsino-like). By the $V-A$ current
structure of the weak interaction, neutrinos tend to be emitted in the
opposite (same) direction to the $\tau$ direction if $\tau$ is
right-handed (left-handed), and that makes the edge sharper
(broader)~\cite{Bullock:1992yt}.
With this structure, we can expect that the $M_{\tilde\tau \tau}$
distribution reconstructed with $\tau$-jet four-momentum shows sharp
edges at three neutralino masses although the Higgsino edges become
slightly weaker.

In the right panel of Fig.~\ref{fig:ejetau} we plotted the same quantity,
$E_{\tau\mbox{-}{\rm jet}}/E_{\tau}$, from all the neutralino decays.
The overall shape, monotonically increasing function and has a sharp
edge at $E_{\tau\mbox{-}{\rm jet}}/E_{\tau}=1$, can be understood from
the distribution of $E_{\tau\mbox{-}{\rm jet}}/E_{\tau}$ in the two-body
decays of $\tau$,
\begin{eqnarray}
\tau\to \pi\nu \ (11\%),
\qquad
\tau\to \rho\nu \ (26\%),
\qquad
\tau\to a_{1}\nu \ (18\%).
\end{eqnarray}
The $\rho$ and $a_{1}$ mesons subsequently decay into two pions and
three pions, respectively.
Here the percentages of each mode denote the branching ratios.
The branching ratio of the leptonic modes are 35\%, and the other 10\%
comes from more than five-body decay modes or the modes with $K$ mesons.
When we ignore the width of the mesons, the energy fraction $E_{\rm
meson}/E_{\tau}$ from each decay distributes uniformly between ($m_{\rm
meson}^{2}/m_{\tau}^{2}$, 1) in the relativistic limit of $\tau$
($E_\tau \gg m_\tau$).  In the figure, we show the distributions of
$E_{\rm meson}/E_{\tau}$ as shaded histograms.
The energy fraction in other hadronic modes tends to pile up near the
edge because of the kinematics of the many body final state.
The distribution of $E_{\rm meson}/E_{\tau}$ well resembles the
properties of the distribution of $E_{\tau\mbox{-}{\rm jet}}/E_{\tau}$.
The thresholds at each meson mass are smeared by the effects of their
finite decay widths (see for e.g.,~\cite{Bullock:1992yt}).

It is important to notice that the distribution of $E_{\tau\mbox{-}{\rm
jet}}/E_{\tau}$ has a tail in the unphysical region,
$E_{\tau\mbox{-}{\rm jet}}/E_{\tau}>1$.  These entries come mainly from
the calibration of the $\tau$-jet energy used in the detector
simulation.
Especially, we should note that the distribution is slightly biased
toward the $E_{\tau\mbox{-}{\rm jet}}/E_{\tau}>1$ region. 
The detailed shape of the distribution, of course, depends on the actual
algorithm for the calibration.
We performed a fitting of the distribution around the peak with a
smeared jagged function $f(x)$,
\begin{eqnarray}
\label{eq:fit}
f(x; x_{0},\sigma,C_{1},C_{2}) &=& \int_{-\infty}^{\infty}dx' 
\frac{g(x-x';x_{0})}{\sqrt{2\pi \sigma^{2}} }
\exp\left[-\frac{x'^{2}}{2 \sigma^{2}}  \right],\\
g(x;x_{0}) &=&
\left\{
\begin{array}{cl}
C_{1}\, x,  &(0<x<x_{0}), \\
C_{2}\, x,  &(x_{0}<x), \\
\end{array}
\right.
\end{eqnarray}
with four fitting parameters, the position of the edge $x_{0}$, the
smearing factor $\sigma$, and two slopes $C_{1}$ and $C_{2}$. The
fitting gives the position of the edge $x_{0}$ to be $x_{0}= 1.049\pm
0.003$, about five percent larger than unity.  Since we will identify
the position of the edge in the $M_{\tilde{\tau}\tau}$ distribution as
the neutralino mass, the bias ends up with systematic errors of the mass
measurement toward larger values.  Therefore, in the actual analysis of
the LHC data, we need to understand the shift of the edge location
caused by the calibration of the $\tau$-jets energy.

\begin{figure}[t]
\begin{center}
  \includegraphics[height=7.4cm]{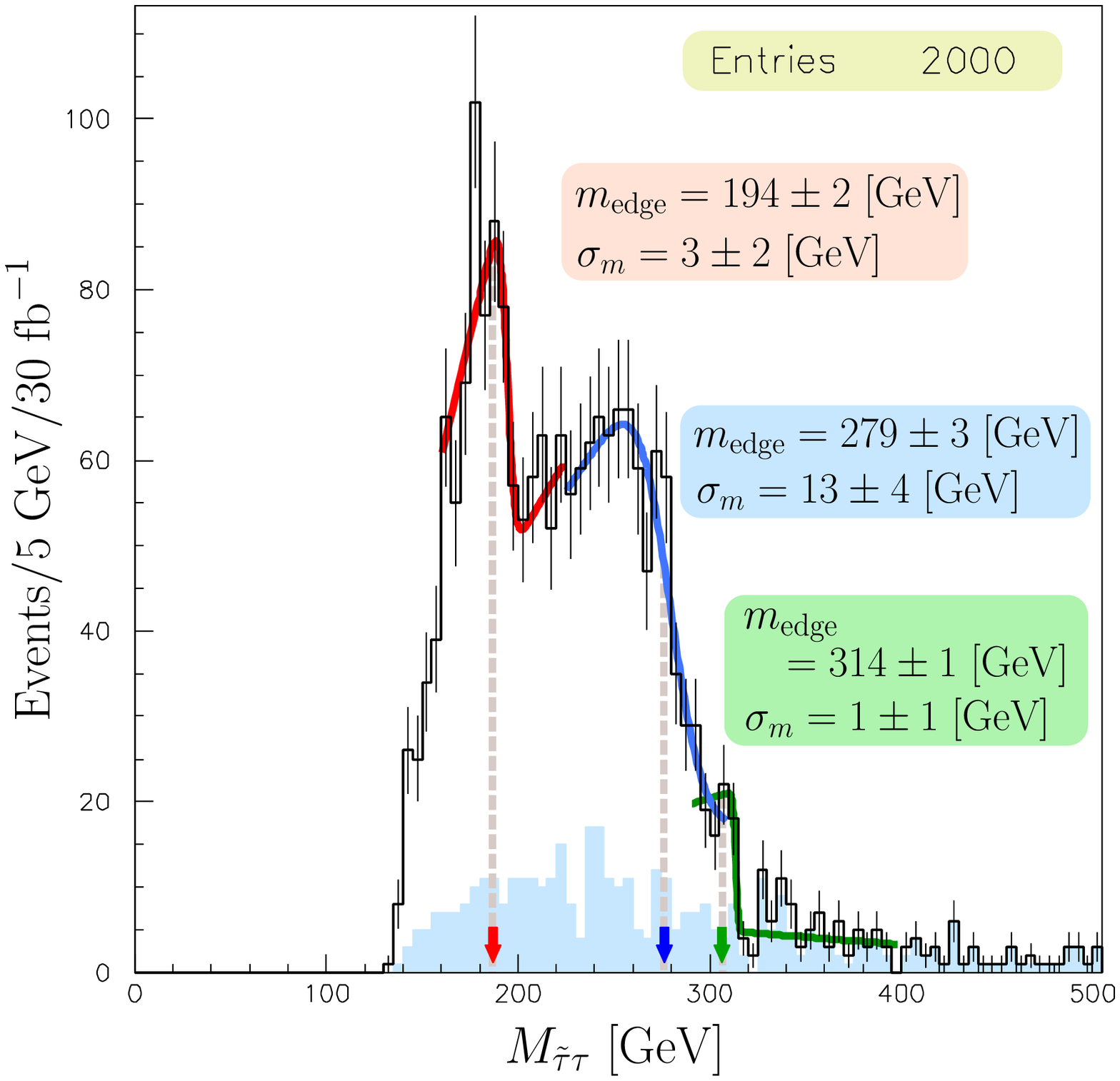}
  \includegraphics[height=7.4cm]{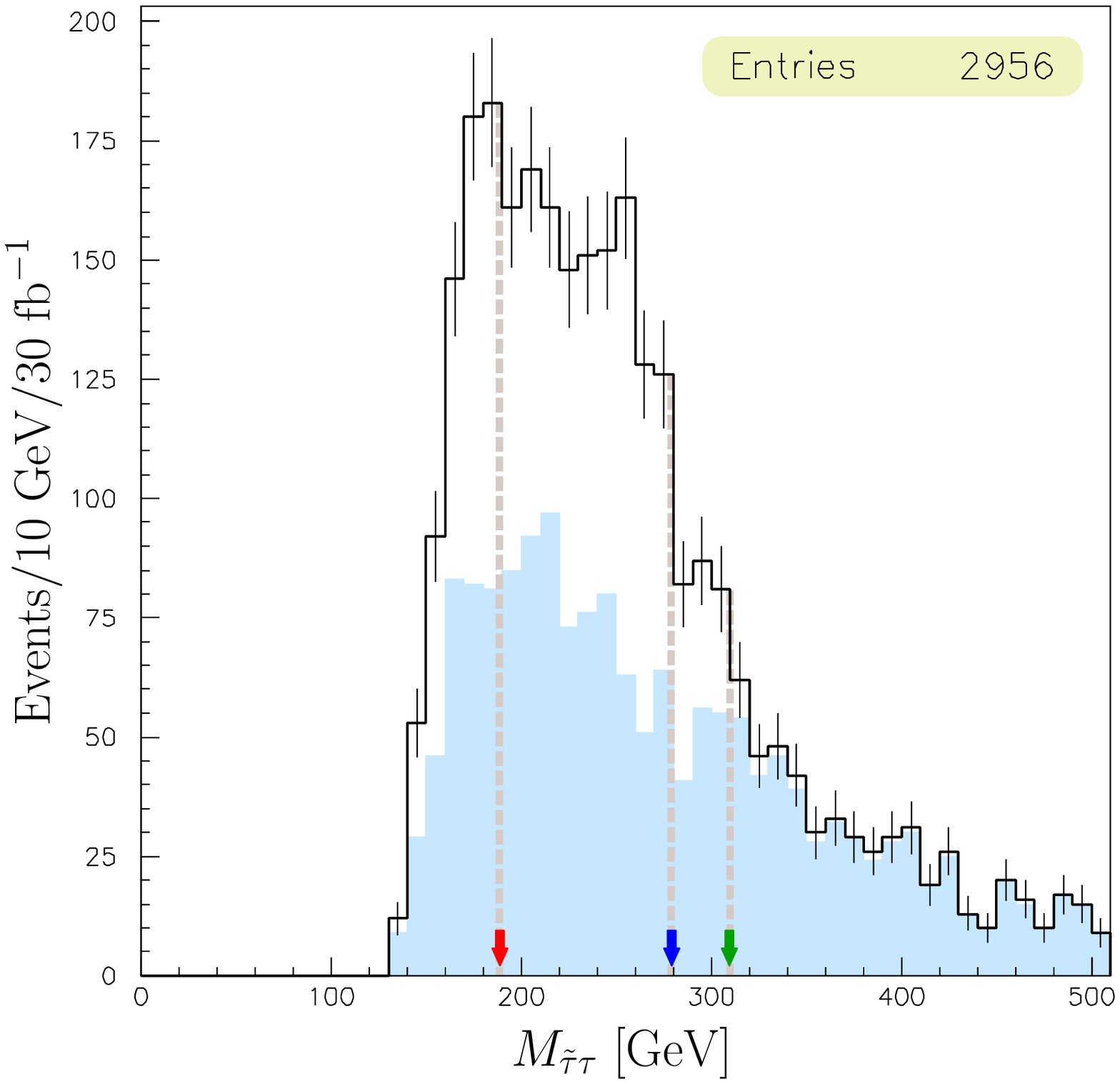}
\end{center}
\caption{Left) The distribution of the lowest invariant mass combination
of $\tilde{\tau}_{1}$ and $\tau$-jet.  The shaded histogram shows the
events with a mis-identified $\tau$-jet which is simulated by assuming a
mis-tagging probability of a non-$\tau$-labelled jet to be 1\%.
The small allows and dashed lines denote the input values of three
neutralino masses.  Three curves are fitting functions of three
endpoints which correspond to the endpoints of $\chi_{1,2,3}^{0}$ from
left to right, respectively. The third endpoint is statistically not
very significant.
Right) The same as the left figure but we assumed the mis-tagging
 probability to be 5\% per a non-$\tau$-labelled jet. 
The endpoints of $\chi_2^0$ and $\chi_3^0$ are visible whereas the
significance of $\chi_1^0$ events are reduced due to the shape of the
background events.  The bin size is 10~GeV in the right figure.}
 \label{fig:mstautau}
\end{figure}

Having understood the edge structure of the $E_{\tau\mbox{-}{\rm
jet}}/E_{\tau}$ distribution, we try to reconstruct the neutralino
masses. 
Fig.~\ref{fig:mstautau} shows the distribution of the smaller invariant
mass out of two possible combinations of $\tilde\tau_{1}$ and the
$\tau$-jet.
We took the mis-tagging probability to be 1\% (left) and 5\% (right).
As we expected, we can clearly see the edge structures in the left
panel. We can determine the masses of $\chi_{1,2,3}^{0}$ from the
location of the edges.  The tail of the distribution for
$M_{\tilde\tau\tau}\gsim 350$\,GeV stems from the mis-tagging of
$\tau$-jets (shaded histograms).
In addition to the background from mis-tagging of $\tau$-jets, the
histogram includes events with a fake stau from muons in supersymmetric
events. The standard model background is assumed to be negligible with
the selection cut discussed before~\cite{Ambrosanio:2000ik}.

In the left figure, we can see the structure that the shape in
Fig.~\ref{fig:parton-tau} is smeared according to the
$E_{\tau\mbox{-}{\rm jet}}/E_{\tau}$ distribution in
Fig.~\ref{fig:ejetau}.
The edge structure of the third neutralino $\chi_3^0$ is not very clear
with the bin size of 5~GeV.
In order to extract the masses of $\chi_{1,2,3}^{0}$, we
have fitted the each endpoints with the smeared jagged function $f(x -
m_{\tilde \tau_1};
m_{\rm edge} - m_{\tilde \tau_1}, \sigma_{m}, C_{1},C_{2})$ in
Eq.~(\ref{eq:fit}) plus a constant ($a_0$) which represents the background
events around the edges.
The results of the fitting are given
in Table.~\ref{tab:fit}.

By taking into account the above physics as well as systematic
uncertainties such as dependence on the calibration algorithm for the
$\tau$-jet momentum, we conclude that the masses of first two (possibly
three) neutralinos can be measured with an accuracy of, at least, about
5\% level.

If we use a loose strategy for the identification of $\tau$, the
background will significantly affects the edge structure. In the right
panel of Fig.~\ref{fig:mstautau}, we have used the mis-tagging
probability to be 5\% per a non-$\tau$-labelled jet.
The edge structure is not significant for $\chi_1^0$ whereas we can see
the edges of $\chi_2^0$ and $\chi_3^0$. (We changed the bin size to
10~GeV in this figure.) This situation will improve when we use a looser
cut on $p_T$ of $\tau$-jets such as $p_T > 20$~GeV.
A similar accuracy for the neutralino mass measurement is possible even
in that case.

\begin{table}[t]
\begin{center}
\begin{tabular}{|c|c|c|c|c|c|c|} 
\hline
      & $m_{\rm edge}$ [GeV] & $\sigma_{m}$ [GeV]  & $C_{1}$ & $C_{2}$ & $a_0$
 & $m_{\chi_{0}}^{\rm input}$ [GeV]\\
 \hline
 ${\chi_{1}^{0}}$        & $194\pm 2$      & $3 \pm 2$  & $0.93\pm 0.35$&
 $0.37\pm 0.35$ & $19 \pm 20$ & 187 \\
 ${\chi_{2}^{0}}$        & $279\pm 3$      & $13 \pm 4 $  & $0.33 \pm
 0.14$ & $-0.014\pm 0.086$ & $20 \pm 17$ & 276\\
 ${\chi_{3}^{0}}$        & $314\pm 1$      & $1 \pm 1$  & $0.066  \pm
 0.027$ & $-0.018\pm0.018$ & $8.3 \pm 4.5$ & 307\\
\hline
\end{tabular}
\end{center}
\caption{The fitting parameters of the function $f(x-m_{\tilde
 \tau_1};m_{\rm edge} - m_{\tilde \tau_1}, \sigma_{m},C_{1},C_{2})$ in
 Eq.~(\ref{eq:fit}) plus a constant $a_0$ around each edge.  The final
 column shows the actual masses of ${\chi^{0}_{1,2,3}}$.  }
 \label{tab:fit}
\end{table}


\subsection{Parameter determination}
\begin{figure}[t]
\begin{center}
  \includegraphics[height=7.6cm]{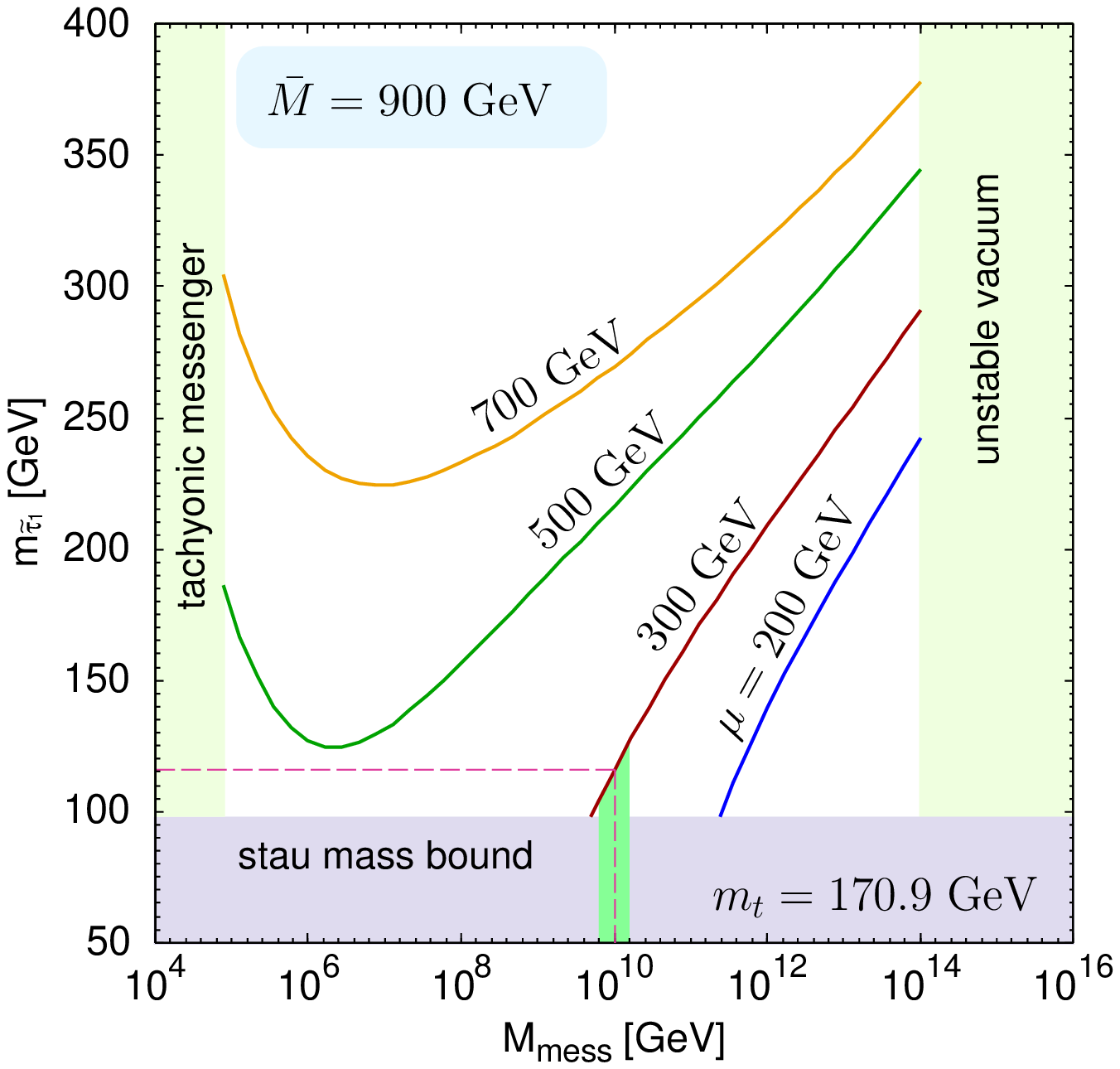}
  \includegraphics[height=7.6cm]{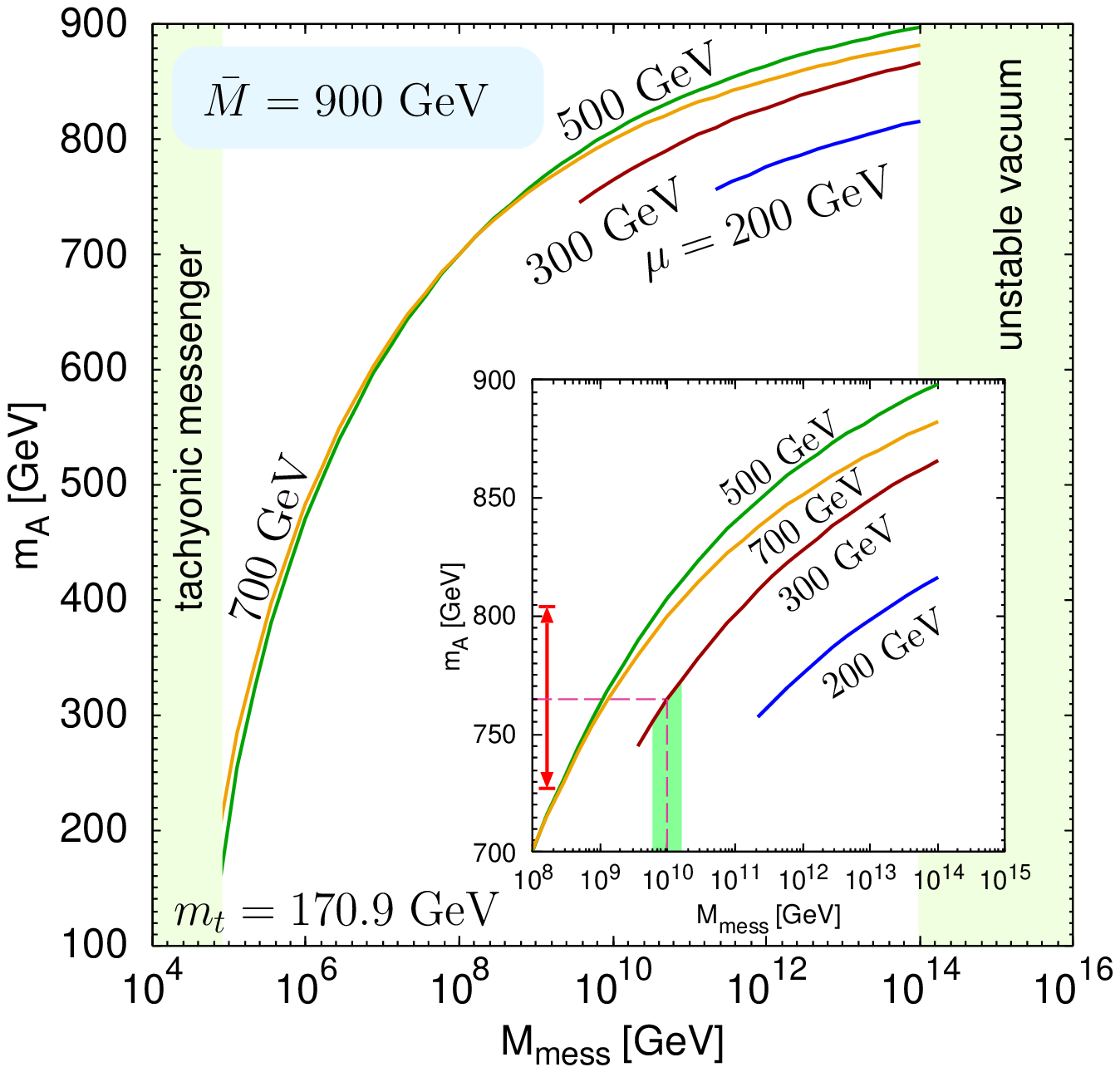}
\end{center}
\caption{Left) The stau mass $m_{\tilde \tau_1}$ as a function of
 $M_{\rm mess}$ for four values of the $\mu$-parameter.  The overall
 scale is set for $\bar{M}=900$\,GeV.  The dashed horizontal line
 corresponds to $\tilde\tau_{1}$ mass, $m_{\tilde\tau_{1}}= 116$\,GeV,
 at the benchmark point.  The thick vertical line denotes the value of
 $M_{\rm mess}$ determined by assuming 5\% precisions of $\mu$ and $\bar
 M$.  Right) The pseudo-scalar Higgs boson mass $m_A$ as a function of
 $M_{\rm mess}$ for four values of the $\mu$-parameter.  The overall
 scale is set for $\bar{M}=900$\,GeV.  The dashed horizontal line
 corresponds to the prediction of $m_{A}$ for $M_{\rm mess} =
 10^{10}$\,GeV.  The thick vertical line denotes the value of $M_{\rm
 mess}$ determined from the stau mass measurement (see the left panel).
 The arrow on the $m_{A}$ axes denotes the error of the prediction
 including the error $\Delta \bar M$.  } \label{fig:stau-mA}
\end{figure}

It is straightforward to determine the model parameters $(\mu, M_{\rm
mess}, \bar M)$ from the measurement of $m_{\tilde \tau_1}$,
$m_{\chi_1^0}$, and $m_{\chi_2^0}$. Performing $\chi^2$ analysis would
either give best fit values of these parameters or exclude the model.

In most cases, a simpler analysis than the global fit is possible.
First, by assuming that the model is correct, we can find that one of
the two neutralinos we measured in the previous section should be
Higgsino-like since their masses deviate from a GUT relation between
$M_{1}$ and $M_{2}$.  
Secondly, we can neglect the $\tan \beta$ dependence in the neutralino
masses. With a large value of $\tan \beta$ (see Fig.~\ref{fig:tanb}),
corrections are of $O(1/\tan \beta)$.
Thus, the neutralino masses depend merely on the messenger scale $M_{\rm
mess}$.
The parameters $\mu$ and $\bar M$ can be determined at the level of 5\%
from the measurement of two leading neutralino masses.  If we can also
measure the mass of $\chi_{3}^{0}$, we can check the consistency of the
GUT relation between $M_{1}$ and $M_{2}$ from the mass splitting between
$\chi_{2}^{0}$ and $\chi_{3}^{0}$, which provides a non-trivial check of
GUT theories.

We can then determine $M_{\rm mess}$ from the measured stau mass. We
demonstrate in the left panel of Fig.~\ref{fig:stau-mA} the
determination of the messenger scale $M_{\rm mess}$.
Once we know the value of $\mu$ and $\bar M$, $m_{\tilde \tau_1}$ can be
calculated as a function of $M_{\rm mess}$.
Since we measure $m_{\tilde{\tau}_{1}}$ at a few permille level, and
$\mu$ and $\bar M$ at 5\% level, we can read off the corresponding value
of $M_{\rm mess}$ from the figure. We find that the exponent of $M_{\rm
mess}$ is determined with an accuracy of $\pm 0.2$.
Therefore, at the benchmark point, the parameters are determined with
the precision of,
\begin{eqnarray}
\Delta\mu \sim 20\,{\rm GeV}, \\
\Delta\bar M \sim 50 \, {\rm GeV}, \\
\Delta \log_{10} M_{{\rm mess}} \sim 0.2.
\end{eqnarray}

Finally, once all the parameters are determined, we can make a
prediction of any other physical quantities.  The simplest test is the
peak location of the $M_{\rm eff}$ distribution which depends on the
squark masses~\cite{Baer:1995nq,Hinchliffe:1996iu,Tovey:2000wk}.
As a more non-trivial example, we show a prediction of the mass of the
pseudo-scalar Higgs boson, $m_{A}$, in the right panel of
Fig.~\ref{fig:stau-mA}.  We can predict a value of $m_{A}$ with an
uncertainty of 5\% level:
\begin{eqnarray}
\Delta m_{A} \sim 40 \,{\rm GeV},
\end{eqnarray}
around $m_{A}=765$\,GeV.  At the LHC, the heavy Higgs boson $H^0/A^0$
will be discovered up to $m_{H/A}\simeq 800$\,GeV for $\tan\beta\simeq
40$ assuming and integrated luminosity 30
fb$^{-1}$\,\cite{Cavalli:2002vs}.  Therefore, we can perform a
non-trivial check of the model from the $m_{A}$ measurement.

\section{Conclusions}

\setcounter{footnote}{0}

In studies of LHC signals of supersymmetric theories, setting the model
parameters is the first non-trivial task which needs to be done. At the
LHC experiments, the main production process of the supersymmetric
particles is a pair production of colored objects, such as a pair
production of gluinos and squarks. It is thus essential to know how
these particles decay. Also, since there are many kinds of particles, it
is often the case that a measurement of supersymmetric parameters
suffers from background processes which also come from supersymmetric
events. In order to estimate the amount of the background, we need to
set all of the parameters in the Lagrangian.

There are, on the other hand, over 100 parameters in the Lagrangian of
the MSSM. It is practically impossible to study every point in the 100
dimensional parameter space. Therefore, parametrizations such as the
mSUGRA model and the ``gauge mediation'' model (the gauginos and scalar
masses from the formula of the gauge
mediation~\cite{Dine:1993yw,Dine:1994vc,Dine:1995ag} and $\mu$ and $B$
parameters as free parameters) have been proposed and used as standard
benchmark models. The number of the parameters is reduced to be a few in
these models.

As a purpose of the study of generic signatures of the supersymmetric
models and for development of methods to extract physical quantities,
these parametrizations have played a significant role in studies of
collider
physics~\cite{Baer:1995nq,Hinchliffe:1996iu,Dimopoulos:1996yq,Hinchliffe:1998ys}.
However, it is dangerous to rely too much on these
parametrizations. Interesting parameter regions in the MSSM can be
precluded by assumptions made without theoretical motivations.
Remember that they are simply convenient parametrizations of the more
than 100 unknown parameters.

The sweet spot scenario we have presented provides an example of a
simple parametrization of the MSSM (by only three parameters), and it is
theoretically supported. It is the first example of such a simple
parametrization which has in background a well-defined closed framework
of the supersymmetry breaking and mediation with no
phenomenological/cosmological problems.

There are new features in the collider signatures. For example, a
relatively light Higgsino is preferred and that makes the decays of
gluino into third generation quarks to be the dominant channel. Many
numbers of $b$-jets will show up in each supersymmetric event. Also, the
light stau is predicted if the Higgsinos are light. If it is the NLSP,
two charged tracks left by escaping staus and $\tau$-jets from the
$\chi^0 \to \tilde \tau \tau$ decays can be used to reconstruct the
three parameters in the Lagrangian as we have demonstrated. The model
confirmation/exclusion is possible by measuring any other quantities
such as the mass of the pseudo-scalar Higgs boson.

\section*{Acknowledgments}

This work was supported by the U.S. Department of Energy under contract
number DE-AC02-76SF00515.



\begin{thebibliography}{0}
\bibitem{Chamseddine:1982jx}
  A.~H.~Chamseddine, R.~Arnowitt and P.~Nath,
  ``Locally Supersymmetric Grand Unification,''
  Phys.\ Rev.\ Lett.\  {\bf 49}, 970 (1982);
  R.~Barbieri, S.~Ferrara and C.~A.~Savoy,
  ``Gauge Models with Spontaneously Broken Local Supersymmetry,''
  Phys.\ Lett.\  B {\bf 119}, 343 (1982).

\bibitem{Hall:1983iz}
  L.~J.~Hall, J.~D.~Lykken and S.~Weinberg,
  ``Supergravity as the Messenger of Supersymmetry Breaking,''
  Phys.\ Rev.\  D {\bf 27} (1983) 2359.

\bibitem{Dine:1981za}
  M.~Dine, W.~Fischler and M.~Srednicki,
  ``Supersymmetric Technicolor,''
  Nucl.\ Phys.\ B {\bf 189}, 575 (1981);
  S.~Dimopoulos and S.~Raby,
  ``Supercolor,''
  Nucl.\ Phys.\ B {\bf 192}, 353 (1981);
  M.~Dine and W.~Fischler,
  ``A Phenomenological Model of Particle Physics Based on Supersymmetry,''
  Phys.\ Lett.\ B {\bf 110}, 227 (1982);
  ``A Supersymmetric Gut,''
  Nucl.\ Phys.\ B {\bf 204}, 346 (1982);
  C.~R.~Nappi and B.~A.~Ovrut,
  ``Supersymmetric Extension of the SU(3) $\times$ SU(2) $\times$ U(1) Model,''
  Phys.\ Lett.\ B {\bf 113}, 175 (1982);
  L.~Alvarez-Gaume, M.~Claudson and M.~B.~Wise,
  ``Low-Energy Supersymmetry,''
  Nucl.\ Phys.\ B {\bf 207}, 96 (1982);
  S.~Dimopoulos and S.~Raby,
  ``Geometric Hierarchy,''
  Nucl.\ Phys.\ B {\bf 219}, 479 (1983).


\bibitem{Dine:1993yw}
  M.~Dine and A.~E.~Nelson,
   ``Dynamical supersymmetry breaking at low-energies,''
  %
  Phys.\ Rev.\ D {\bf 48}, 1277 (1993)
  [arXiv:hep-ph/9303230].

\bibitem{Dine:1994vc}
  M.~Dine, A.~E.~Nelson and Y.~Shirman,
   ``Low-energy dynamical supersymmetry breaking simplified,''
  %
  Phys.\ Rev.\ D {\bf 51}, 1362 (1995)
  [arXiv:hep-ph/9408384].

\bibitem{Dine:1995ag}
  M.~Dine, A.~E.~Nelson, Y.~Nir and Y.~Shirman,
   ``New tools for low-energy dynamical supersymmetry breaking,''
  %
  Phys.\ Rev.\ D {\bf 53}, 2658 (1996)
  [arXiv:hep-ph/9507378].

\bibitem{Giudice:1997ni}
  G.~F.~Giudice and R.~Rattazzi,
  ``Extracting supersymmetry-breaking effects from wave-function
  renormalization,''
  Nucl.\ Phys.\  B {\bf 511}, 25 (1998)
  [arXiv:hep-ph/9706540].

\bibitem{Randall:1998uk}
  L.~Randall and R.~Sundrum,
  ``Out of this world supersymmetry breaking,''
  Nucl.\ Phys.\ B {\bf 557}, 79 (1999)
  [arXiv:hep-th/9810155];
  G.~F.~Giudice, M.~A.~Luty, H.~Murayama and R.~Rattazzi,
  ``Gaugino mass without singlets,''
  JHEP {\bf 9812}, 027 (1998)
  [arXiv:hep-ph/9810442].

\bibitem{Coughlan:1983ci}
  G.~D.~Coughlan, W.~Fischler, E.~W.~Kolb, S.~Raby and G.~G.~Ross,
  ``Cosmological Problems For The Polonyi Potential,''
  Phys.\ Lett.\ B {\bf 131}, 59 (1983).

\bibitem{O'Raifeartaigh:1975pr}
  L.~O'Raifeartaigh,
  ``Spontaneous Symmetry Breaking for Chiral Scalar Superfields,''
  Nucl.\ Phys.\  B {\bf 96}, 331 (1975).

\bibitem{Giudice:1988yz}
  G.~F.~Giudice and A.~Masiero,
  ``A Natural Solution to the $\mu$ Problem in Supergravity Theories,''
  Phys.\ Lett.\  B {\bf 206}, 480 (1988).

\bibitem{Gabbiani:1996hi}
  F.~Gabbiani, E.~Gabrielli, A.~Masiero and L.~Silvestrini,
  ``A complete analysis of FCNC and CP constraints in general SUSY extensions
  of the standard model,''
  Nucl.\ Phys.\  B {\bf 477}, 321 (1996)
  [arXiv:hep-ph/9604387].

\bibitem{Banks:1993en}
  T.~Banks, D.~B.~Kaplan and A.~E.~Nelson,
  ``Cosmological implications of dynamical supersymmetry breaking,''
  Phys.\ Rev.\  D {\bf 49}, 779 (1994)
  [arXiv:hep-ph/9308292].

\bibitem{Dine:1983ys}
  M.~Dine, W.~Fischler and D.~Nemeschansky,
  ``Solution of the Entropy Crisis of Supersymmetric Theories,''
  Phys.\ Lett.\  B {\bf 136}, 169 (1984).

\bibitem{Joichi:1994ce}
  I.~Joichi and M.~Yamaguchi,
  ``Heavy Polonyi field as a solution of the Polonyi problem,''
  Phys.\ Lett.\  B {\bf 342}, 111 (1995)
  [arXiv:hep-ph/9409266].

\bibitem{Ibe:2006am}
  M.~Ibe, Y.~Shinbara and T.~T.~Yanagida,
  ``The Polonyi problem and upper bound on inflation scale in supergravity,''
  Phys.\ Lett.\  B {\bf 639}, 534 (2006)
  [arXiv:hep-ph/0605252].

\bibitem{Ibe:2006rc}
  M.~Ibe and R.~Kitano,
  ``Gauge mediation in supergravity and gravitino dark matter,''
  Phys.\ Rev.\  D {\bf 75}, 055003 (2007)
  [arXiv:hep-ph/0611111].

\bibitem{Dvali:1996cu}
  G.~R.~Dvali, G.~F.~Giudice and A.~Pomarol,
  ``The $\mu$-Problem in Theories with Gauge-Mediated Supersymmetry Breaking,''
  Nucl.\ Phys.\  B {\bf 478}, 31 (1996)
  [arXiv:hep-ph/9603238].

\bibitem{Giudice:1998bp}
  G.~F.~Giudice and R.~Rattazzi,
  ``Theories with gauge-mediated supersymmetry breaking,''
  Phys.\ Rept.\  {\bf 322}, 419 (1999)
  [arXiv:hep-ph/9801271].

\bibitem{deGouvea:1997cx}
  A.~de Gouvea, A.~Friedland and H.~Murayama,
  ``Next-to-minimal supersymmetric standard model with the gauge mediation  of
  supersymmetry breaking,''
  Phys.\ Rev.\  D {\bf 57}, 5676 (1998)
  [arXiv:hep-ph/9711264].

\bibitem{Nelson:1993nf}
  A.~E.~Nelson and N.~Seiberg,
  ``R symmetry breaking versus supersymmetry breaking,''
  Nucl.\ Phys.\  B {\bf 416}, 46 (1994)
  [arXiv:hep-ph/9309299].

\bibitem{Kitano:2006wz}
  R.~Kitano,
  ``Gravitational gauge mediation,''
  Phys.\ Lett.\  B {\bf 641}, 203 (2006)
  [arXiv:hep-ph/0607090].

\bibitem{Kitano:2006xg}
  R.~Kitano, H.~Ooguri and Y.~Ookouchi,
  ``Direct mediation of meta-stable supersymmetry breaking,''
  Phys.\ Rev.\  D {\bf 75}, 045022 (2007)
  [arXiv:hep-ph/0612139].

\bibitem{Murayama:2006yf}
  H.~Murayama and Y.~Nomura,
  ``Gauge mediation simplified,''
  Phys.\ Rev.\ Lett.\  {\bf 98}, 151803 (2007)
  [arXiv:hep-ph/0612186];
  H.~Murayama and Y.~Nomura,
  ``Simple scheme for gauge mediation,''
  arXiv:hep-ph/0701231.

\bibitem{Aharony:2006my}
  O.~Aharony and N.~Seiberg,
  ``Naturalized and simplified gauge mediation,''
  JHEP {\bf 0702}, 054 (2007)
  [arXiv:hep-ph/0612308].

\bibitem{Intriligator:2006dd}
  K.~Intriligator, N.~Seiberg and D.~Shih,
  ``Dynamical SUSY breaking in meta-stable vacua,''
  JHEP {\bf 0604}, 021 (2006)
  [arXiv:hep-th/0602239].

\bibitem{Dine:2006gm}
  M.~Dine, J.~L.~Feng and E.~Silverstein,
  ``Retrofitting O'Raifeartaigh models with dynamical scales,''
  Phys.\ Rev.\  D {\bf 74}, 095012 (2006)
  [arXiv:hep-th/0608159].

\bibitem{Dine:2006xt}
  M.~Dine and J.~Mason,
  ``Gauge mediation in metastable vacua,''
  arXiv:hep-ph/0611312.

\bibitem{Csaki:2006wi}
  C.~Csaki, Y.~Shirman and J.~Terning,
  ``A simple model of low-scale direct gauge mediation,''
  arXiv:hep-ph/0612241.

\bibitem{Ellwanger:1983mg}
  U.~Ellwanger,
  ``Nonrenormalizable Interactions From Supergravity, Quantum Corrections And
  Effective Low-Energy Theories,''
  Phys.\ Lett.\  B {\bf 133} (1983) 187.

\bibitem{Bagger:1993ji}
  J.~Bagger and E.~Poppitz,
  ``Destabilizing divergences in supergravity coupled supersymmetric
  theories,''
  Phys.\ Rev.\ Lett.\  {\bf 71}, 2380 (1993)
  [arXiv:hep-ph/9307317].

\bibitem{Chacko:1999am}
  Z.~Chacko, M.~A.~Luty, I.~Maksymyk and E.~Ponton,
  ``Realistic anomaly-mediated supersymmetry breaking,''
  JHEP {\bf 0004}, 001 (2000)
  [arXiv:hep-ph/9905390].

\bibitem{Ibe:2004gh}
  M.~Ibe, R.~Kitano and H.~Murayama,
  ``A viable supersymmetric model with UV insensitive anomaly mediation,''
  Phys.\ Rev.\  D {\bf 71}, 075003 (2005)
  [arXiv:hep-ph/0412200].

\bibitem{Kaplan:1999ac}
  D.~E.~Kaplan, G.~D.~Kribs and M.~Schmaltz,
  ``Supersymmetry breaking through transparent extra dimensions,''
  Phys.\ Rev.\  D {\bf 62}, 035010 (2000)
  [arXiv:hep-ph/9911293];
  Z.~Chacko, M.~A.~Luty, A.~E.~Nelson and E.~Ponton,
  ``Gaugino mediated supersymmetry breaking,''
  JHEP {\bf 0001}, 003 (2000)
  [arXiv:hep-ph/9911323].

\bibitem{Pomarol:1999ie}
  A.~Pomarol and R.~Rattazzi,
  ``Sparticle masses from the superconformal anomaly,''
  JHEP {\bf 9905}, 013 (1999)
  [arXiv:hep-ph/9903448].

\bibitem{Chacko:2001jt}
  Z.~Chacko and M.~A.~Luty,
  ``Realistic anomaly mediation with bulk gauge fields,''
  JHEP {\bf 0205}, 047 (2002)
  [arXiv:hep-ph/0112172];
  R.~Sundrum,
  ``'Gaugomaly' mediated SUSY breaking and conformal sequestering,''
  Phys.\ Rev.\  D {\bf 71}, 085003 (2005)
  [arXiv:hep-th/0406012].

\bibitem{Kitano:2006wm}
  R.~Kitano,
  ``Dynamical GUT breaking and $\mu$-term driven supersymmetry breaking,''
  Phys.\ Rev.\  D {\bf 74}, 115002 (2006)
  [arXiv:hep-ph/0606129].

\bibitem{Ritt:2006cg}
  S.~Ritt  [MEG Collaboration],
  ``Status of the MEG Expriment $\mu \to e \gamma$,''
  Nucl.\ Phys.\ Proc.\ Suppl.\  {\bf 162} (2006) 279.

\bibitem{Kuno:2005mm}
  Y.~Kuno,
  ``PRISM/PRIME,''
  Nucl.\ Phys.\ Proc.\ Suppl.\  {\bf 149}, 376 (2005).

\bibitem{Hall:1985dx}
  L.~J.~Hall, V.~A.~Kostelecky and S.~Raby,
  ``New Flavor Violations In Supergravity Models,''
  Nucl.\ Phys.\  B {\bf 267}, 415 (1986).

\bibitem{Borzumati:1986qx}
  F.~Borzumati and A.~Masiero,
  ``Large Muon And Electron Number Violations In Supergravity Theories,''
  Phys.\ Rev.\ Lett.\  {\bf 57}, 961 (1986).

\bibitem{Pospelov:2006sc}
  M.~Pospelov,
  ``Particle physics catalysis of thermal big bang nucleosynthesis,''
  arXiv:hep-ph/0605215;

\bibitem{Kohri:2006cn}
  K.~Kohri and F.~Takayama,
  ``Big bang nucleosynthesis with long lived charged massive particles,''
  arXiv:hep-ph/0605243;

\bibitem{Cyburt:2006uv}
  R.~H.~Cyburt, J.~Ellis, B.~D.~Fields, K.~A.~Olive and V.~C.~Spanos,
  ``Bound-state effects on light-element abundances in gravitino dark  matter
  scenarios,''
  JCAP {\bf 0611}, 014 (2006)
  [arXiv:astro-ph/0608562].

\bibitem{Hamaguchi:2007mp}
  K.~Hamaguchi, T.~Hatsuda, M.~Kamimura, Y.~Kino and T.~T.~Yanagida,
  ``Stau-catalyzed Li-6 production in big-bang nucleosynthesis,''
  arXiv:hep-ph/0702274;
  M.~Kawasaki, K.~Kohri and T.~Moroi,
  ``Big-bang nucleosynthesis with long-lived charged slepton,''
  arXiv:hep-ph/0703122.

\bibitem{Kaplinghat:2006qr}
  M.~Kaplinghat and A.~Rajaraman,
  ``Big bang nucleosynthesis with bound states of long-lived charged
  particles,''
  Phys.\ Rev.\  D {\bf 74}, 103004 (2006)
  [arXiv:astro-ph/0606209];
  C.~Bird, K.~Koopmans and M.~Pospelov,
  ``Primordial Lithium Abundance in Catalyzed Big Bang Nucleosynthesis,''
  arXiv:hep-ph/0703096;
  T.~Jittoh, K.~Kohri, M.~Koike, J.~Sato, T.~Shimomura and M.~Yamanaka,
  ``Possible solution to the $^7$Li problem by the long lived stau,''
  arXiv:0704.2914 [hep-ph].

\bibitem{Goto:1998qg}
  T.~Goto and T.~Nihei,
  ``Effect of RRRR dimension 5 operator on the proton decay in the minimal
  SU(5) SUGRA GUT model,''
  Phys.\ Rev.\  D {\bf 59}, 115009 (1999)
  [arXiv:hep-ph/9808255].

\bibitem{Intriligator:1995id}
  K.~A.~Intriligator and N.~Seiberg,
  ``Duality, monopoles, dyons, confinement and oblique confinement in
  supersymmetric $SO(N_c)$ gauge theories,''
  Nucl.\ Phys.\  B {\bf 444}, 125 (1995)
  [arXiv:hep-th/9503179].

\bibitem{Seiberg:1994pq}
  N.~Seiberg,
  ``Electric - magnetic duality in supersymmetric nonAbelian gauge theories,''
  Nucl.\ Phys.\  B {\bf 435}, 129 (1995)
  [arXiv:hep-th/9411149].

\bibitem{Luty:1997fk}
  M.~A.~Luty,
  ``Naive dimensional analysis and supersymmetry,''
  Phys.\ Rev.\  D {\bf 57}, 1531 (1998)
  [arXiv:hep-ph/9706235].

\bibitem{Hotta:1996qb}
  T.~Hotta, K.~I.~Izawa and T.~Yanagida,
  ``Natural Unification with a Supersymmetric SO(10)$_{GUT}$ x SO(6)$_H$ Gauge
  Theory,''
  Phys.\ Rev.\  D {\bf 54}, 6970 (1996)
  [arXiv:hep-ph/9602439].

\bibitem{Barate:2003sz}
  R.~Barate {\it et al.}  [LEP Working Group for Higgs boson searches],
  ``Search for the standard model Higgs boson at LEP,''
  Phys.\ Lett.\  B {\bf 565}, 61 (2003)
  [arXiv:hep-ex/0306033];
    [LEP Higgs Working Group],
  ``Searches for the neutral Higgs bosons of the MSSM: Preliminary combined
  results using LEP data collected at energies up to 209-GeV,''
  arXiv:hep-ex/0107030.

\bibitem{Okada:1990vk}
  Y.~Okada, M.~Yamaguchi and T.~Yanagida,
  ``Upper bound of the lightest Higgs boson mass in the minimal supersymmetric
  standard model,''
  Prog.\ Theor.\ Phys.\  {\bf 85}, 1 (1991);
  J.~R.~Ellis, G.~Ridolfi and F.~Zwirner,
  ``Radiative corrections to the masses of supersymmetric Higgs bosons,''
  Phys.\ Lett.\  B {\bf 257}, 83 (1991);
  H.~E.~Haber and R.~Hempfling,
  ``Can the mass of the lightest Higgs boson of the minimal supersymmetric
  model be larger than m(Z)?,''
  Phys.\ Rev.\ Lett.\  {\bf 66}, 1815 (1991).

\bibitem{Kitano:2006gv}
  R.~Kitano and Y.~Nomura,
  ``Supersymmetry, naturalness, and signatures at the LHC,''
  Phys.\ Rev.\  D {\bf 73}, 095004 (2006)
  [arXiv:hep-ph/0602096].

\bibitem{Peccei:1977hh}
  R.~D.~Peccei and H.~R.~Quinn,
  ``CP Conservation in the Presence of Instantons,''
  Phys.\ Rev.\ Lett.\  {\bf 38}, 1440 (1977);
  R.~D.~Peccei and H.~R.~Quinn,
  ``Constraints Imposed by CP Conservation in the Presence of Instantons,''
  Phys.\ Rev.\  D {\bf 16}, 1791 (1977).

\bibitem{Goto:1991gq}
  T.~Goto and M.~Yamaguchi,
  ``Is Axino Dark Matter Possible In Supergravity?,''
  Phys.\ Lett.\  B {\bf 276}, 103 (1992).

\bibitem{Rattazzi:1996fb}
  R.~Rattazzi and U.~Sarid,
  ``Large tan$\beta$ in gauge-mediated SUSY-breaking models,''
  Nucl.\ Phys.\  B {\bf 501}, 297 (1997)
  [arXiv:hep-ph/9612464].

\bibitem{Skands:2003cj}
  P.~Skands {\it et al.},
  ``SUSY Les Houches accord: Interfacing SUSY spectrum calculators, decay
  packages, and event generators,''
  JHEP {\bf 0407}, 036 (2004)
  [arXiv:hep-ph/0311123].

\bibitem{unknown:2007bx}
    [CDF Collaboration],
  ``A combination of CDF and D0 results on the mass of the top quark,''
  arXiv:hep-ex/0703034.

\bibitem{Abbiendi:2003yd}
  G.~Abbiendi {\it et al.}  [OPAL Collaboration],
  ``Search for stable and long-lived massive charged particles in $e^+$$e^-$
  collisions at $\sqrt s$ = 130~GeV to 209~GeV,''
  Phys.\ Lett.\  B {\bf 572}, 8 (2003)
  [arXiv:hep-ex/0305031].



















\bibitem{Drees:1990yw}
  M.~Drees and X.~Tata,
``Signals for heavy exotics at hadron colliders and supercolliders,''
  Phys.\ Lett.\  B {\bf 252}, 695 (1990).
\bibitem{Dimopoulos:1996yq}
  S.~Dimopoulos, S.~D.~Thomas and J.~D.~Wells,
``Sparticle spectroscopy and electroweak symmetry breaking with
gauge-mediated supersymmetry breaking,''
  Nucl.\ Phys.\  B {\bf 488}, 39 (1997)
  [arXiv:hep-ph/9609434].
\bibitem{Ambrosanio:1997rv}
  S.~Ambrosanio, G.~D.~Kribs and S.~P.~Martin,
``Signals for gauge-mediated supersymmetry breaking models at the CERN  LEP2
collider,''
  Phys.\ Rev.\  D {\bf 56}, 1761 (1997)
  [arXiv:hep-ph/9703211].
\bibitem{Nisati:1997gb}
  A.~Nisati, S.~Petrarca and G.~Salvini,
``On the possible detection of massive stable exotic particles at the  LHC,''
  Mod.\ Phys.\ Lett.\  A {\bf 12}, 2213 (1997)
  [arXiv:hep-ph/9707376].
\bibitem{Feng:1997zr}
  J.~L.~Feng and T.~Moroi,
``Tevatron signatures of long-lived charged sleptons in gauge-mediated
supersymmetry breaking models,''
  Phys.\ Rev.\  D {\bf 58}, 035001 (1998)
  [arXiv:hep-ph/9712499].
\bibitem{Hinchliffe:1998ys}
  I.~Hinchliffe and F.~E.~Paige,
``Measurements in gauge mediated SUSY breaking models at LHC,''
  Phys.\ Rev.\  D {\bf 60}, 095002 (1999)
  [arXiv:hep-ph/9812233].
\bibitem{Mercadante:2000hw}
  P.~G.~Mercadante, J.~K.~Mizukoshi and H.~Yamamoto,
``Analysis of long-lived slepton NLSP in GMSB model at linear collider,''
  Phys.\ Rev.\  D {\bf 64}, 015005 (2001)
  [arXiv:hep-ph/0010067].
\bibitem{Ambrosanio:2000ik}
  S.~Ambrosanio, B.~Mele, S.~Petrarca, G.~Polesello and A.~Rimoldi,
``Measuring the SUSY breaking scale at the LHC in the slepton NLSP  scenario
of GMSB models,''
  JHEP {\bf 0101}, 014 (2001)
  [arXiv:hep-ph/0010081].
  
\bibitem{Buchmuller:2004rq}
  W.~Buchmuller, K.~Hamaguchi, M.~Ratz and T.~Yanagida,
``Supergravity at colliders,''
  Phys.\ Lett.\  B {\bf 588}, 90 (2004)
  [arXiv:hep-ph/0402179].
\bibitem{Feng:2004mt}
  J.~L.~Feng, S.~Su and F.~Takayama,
``Supergravity with a gravitino LSP,''
  Phys.\ Rev.\  D {\bf 70}, 075019 (2004)
  [arXiv:hep-ph/0404231].
\bibitem{Ellis:2006vu}
  J.~R.~Ellis, A.~R.~Raklev and O.~K.~Oye,
``Gravitino dark matter scenarios with massive metastable charged  sparticles
  at the LHC,''
  JHEP {\bf 0610}, 061 (2006)
  [arXiv:hep-ph/0607261].
\bibitem{Cakir:2007xa}
  O.~Cakir, I.~T.~Cakir, J.~R.~Ellis and Z.~Kirca,
``Measurements of metastable staus at linear colliders,''
  arXiv:hep-ph/0703121.
  
\bibitem{Hamaguchi:2004df}
  K.~Hamaguchi, Y.~Kuno, T.~Nakaya and M.~M.~Nojiri,
``A study of late decaying charged particles at future colliders,''
  Phys.\ Rev.\  D {\bf 70}, 115007 (2004)
  [arXiv:hep-ph/0409248].
\bibitem{Feng:2004yi}
  J.~L.~Feng and B.~T.~Smith,
``Slepton trapping at the Large Hadron and International Linear  Colliders,''
  Phys.\ Rev.\  D {\bf 71}, 015004 (2005)
  [Erratum-ibid.\  D {\bf 71}, 0109904 (2005)]
  [arXiv:hep-ph/0409278].
\bibitem{DeRoeck:2005bw}
  A.~De Roeck, J.~R.~Ellis, F.~Gianotti, F.~Moortgat, K.~A.~Olive and L.~Pape,
``Supersymmetric benchmarks with non-universal scalar masses or gravitino
dark matter,''
  Eur.\ Phys.\ J.\  C {\bf 49}, 1041 (2007)
  [arXiv:hep-ph/0508198].
\bibitem{Martyn:2006as}
  H.~U.~Martyn,
``Detecting metastable staus and gravitinos at the ILC,''
  Eur.\ Phys.\ J.\  C {\bf 48}, 15 (2006)
  [arXiv:hep-ph/0605257].
\bibitem{Hamaguchi:2006vu}
  K.~Hamaguchi, M.~M.~Nojiri and A.~de Roeck,
``Prospects to study a long-lived charged next lightest supersymmetric
particle at the LHC,''
  JHEP {\bf 0703}, 046 (2007)
  [arXiv:hep-ph/0612060].

\bibitem{Eidelman:2004wy}
  S.~Eidelman {\it et al.}  [Particle Data Group],
``Review of particle physics,''
  Phys.\ Lett.\  B {\bf 592}, 1 (2004).

\bibitem{Drees:1991mx}
  M.~Drees and M.~M.~Nojiri,
  ``One Loop Corrections To The Higgs Sector In Minimal Supergravity Models,''
  Phys.\ Rev.\  D {\bf 45}, 2482 (1992).

\bibitem{Heinemeyer:1998yj}
  S.~Heinemeyer, W.~Hollik and G.~Weiglein,
  ``FeynHiggs: A program for the calculation of the masses of the neutral
  CP-even Higgs bosons in the MSSM,''
  Comput.\ Phys.\ Commun.\  {\bf 124}, 76 (2000)
  [arXiv:hep-ph/9812320].

\bibitem{Djouadi:2002ze}
  A.~Djouadi, J.~L.~Kneur and G.~Moultaka,
  ``SuSpect: A Fortran code for the supersymmetric and Higgs particle spectrum
  in the MSSM,''
  Comput.\ Phys.\ Commun.\  {\bf 176}, 426 (2007)
  [arXiv:hep-ph/0211331].

\bibitem{Paige:2003mg}
  F.~E.~Paige, S.~D.~Protopopescu, H.~Baer and X.~Tata,
``ISAJET 7.69: A Monte Carlo event generator for $pp$, $\bar p p$, and
	$e^+ e^-$ reactions,''
  arXiv:hep-ph/0312045.

\bibitem{Hinchliffe:1996iu}
  I.~Hinchliffe, F.~E.~Paige, M.~D.~Shapiro, J.~Soderqvist and W.~Yao,
  ``Precision SUSY measurements at LHC,''
  Phys.\ Rev.\  D {\bf 55}, 5520 (1997)
  [arXiv:hep-ph/9610544];
  H.~Bachacou, I.~Hinchliffe and F.~E.~Paige,
  ``Measurements of masses in SUGRA models at LHC,''
  Phys.\ Rev.\  D {\bf 62}, 015009 (2000)
  [arXiv:hep-ph/9907518].

\bibitem{Corcella:2002jc}
  G.~Corcella {\it et al.},
``HERWIG 6.5 release note,''
  arXiv:hep-ph/0210213.
\bibitem{Martin:1998np}
  A.~D.~Martin, R.~G.~Roberts, W.~J.~Stirling and R.~S.~Thorne,
  ``Scheme dependence, leading order and higher twist studies of MRST
  partons,''
  Phys.\ Lett.\  B {\bf 443}, 301 (1998)
  [arXiv:hep-ph/9808371].

\bibitem{Jadach:1993hs}
  S.~Jadach, Z.~Was, R.~Decker and J.~H.~Kuhn,
``The Tau Decay Library TAUOLA: Version 2.4,''
  Comput.\ Phys.\ Commun.\  {\bf 76}, 361 (1993).

\bibitem{Richter-Was:2002ch}
  E.~Richter-Was,
``AcerDET: A particle level fast simulation and reconstruction package  for
phenomenological studies on high $p_T$ physics at LHC,''
  arXiv:hep-ph/0207355.

\bibitem{tdr:1999fq}
  ATLAS Collaboration,
  ``ATLAS: Detector and physics performance technical design report. Volume
  1,''
  CERN-LHCC-99-14.

\bibitem{tautag}
 S.~Rajagopalan, talk given at the TeV4LHC workshop, BNL, February 2005;
 M.~Heldmann, talk given at the TeV4LHC workshop, Fermilab, October
	2005;
  R.~Arnowitt {\it et al.},
  ``Measuring the $\tilde \tau$ - $\tilde \chi_1^0$ mass difference in co-annihilation
  scenarios at the LHC,''
  arXiv:hep-ph/0608193;
  C.~Galea  [D0 Collaboration],
  ``Tau identification at D0,''
  Acta Phys.\ Polon.\  B {\bf 38}, 769 (2007).


\bibitem{Bullock:1992yt}
  B.~K.~Bullock, K.~Hagiwara and A.~D.~Martin,
  ``Tau Polarization and its Correlations as a Probe of New Physics,''
  Nucl.\ Phys.\  B {\bf 395}, 499 (1993).

\bibitem{Baer:1995nq}
  H.~Baer, C.~h.~Chen, F.~Paige and X.~Tata,
  ``Signals for minimal supergravity at the CERN large hadron collider: Multi -
  jet plus missing energy channel,''
  Phys.\ Rev.\  D {\bf 52}, 2746 (1995)
  [arXiv:hep-ph/9503271];
  H.~Baer, C.~h.~Chen, F.~Paige and X.~Tata,
  ``Signals for Minimal Supergravity at the CERN Large Hadron Collider II:
  Multilepton Channels,''
  Phys.\ Rev.\  D {\bf 53}, 6241 (1996)
  [arXiv:hep-ph/9512383].

\bibitem{Tovey:2000wk}
  D.~R.~Tovey,
  ``Measuring the SUSY mass scale at the LHC,''
  Phys.\ Lett.\  B {\bf 498}, 1 (2001)
  [arXiv:hep-ph/0006276].

\bibitem{Cavalli:2002vs}
  D.~Cavalli {\it et al.},
``The Higgs working group: Summary report,''
  arXiv:hep-ph/0203056.





\end{thebibliography}
\end{document}